\documentclass[letter,12pt]{report}

\usepackage{amsmath,amssymb}             % AMS Math
\usepackage[utf8]{inputenc}
\usepackage[OT1]{fontenc}
\usepackage[left=2.5cm,right=2.5cm,top=2cm,bottom=2cm,includefoot,includehead,headheight=13.6pt]{geometry}
\usepackage{setspace}
\usepackage{lineno}
\usepackage{footmisc}
\usepackage{indentfirst}
\usepackage{siunitx}
\usepackage{lmodern}
\usepackage{bm}
\usepackage{float}% If comment this, figure moves to Page 2
\usepackage{tabu}
\usepackage{multirow}
\usepackage{cite}
\usepackage[toc,page]{appendix}

\usepackage{graphicx,type1cm,eso-pic,color}

\usepackage{color}
\definecolor{linkcol}{rgb}{0,0,0.4} 
\definecolor{citecol}{rgb}{0.5,0,0} 

\usepackage{array}
\newcolumntype{L}[1]{>{\raggedright\let\newline\\\arraybackslash\hspace{0pt}}m{#1}}
\newcolumntype{C}[1]{>{\centering\let\newline\\\arraybackslash\hspace{0pt}}m{#1}}
\newcolumntype{R}[1]{>{\raggedleft\let\newline\\\arraybackslash\hspace{0pt}}m{#1}}

\usepackage{rotating}                    % Sideways of figures & tables
\usepackage{fancyhdr}                    % Fancy Header and Footer

\let\headruleORIG\headrule
\renewcommand{\headrule}{\color{black} \headruleORIG}

\usepackage{colortbl}
\arrayrulecolor{black}

\fancypagestyle{plain}{
  \fancyhead{}
  \fancyfoot{}
  
}

\makeatletter

\def\cleardoublepage{\clearpage\if@twoside \ifodd\c@page\else%
  \hbox{}%
  \thispagestyle{empty}%              % Empty header styles
  \newpage%
  \if@twocolumn\hbox{}\newpage\fi\fi\fi}

\makeatother
 
{%

\hrulefill
\vspace*{0.5cm}%
\end{minipage}
}

{ \begin{list}%
	{$\bullet$}%
	{\setlength{\labelwidth}{25pt}%
	 \setlength{\leftmargin}{30pt}%
	 \setlength{\itemsep}{\parsep}}}%
{ \end{list} }

\renewcommand{\epsilon}{\varepsilon}

\begin{document}

\begin{titlepage}
     \begin{center}
       \vspace*{-1.8cm}
        
       \noindent {\Large \textbf{Jefferson Lab PAC 47} } \\
       \noindent \Huge \textbf{Neutron DVCS Measurements with BONuS12 in 
        CLAS12} \\
%        \vspace*{0.7cm}
     \end{center}
   
\renewcommand{\thefootnote}{\fnsymbol{footnote}}
     \begin{center}

       \vspace*{0.6cm}
        \noindent {M.~Hattawy$^\dagger$\footnote[3]{Contact person: 
        Hattawy@jlab.org}, M.~Amaryan, S.~B\"ultmann, G.~Dodge, 
        N.~Dzbenski, C.~Hyde,  S.~Kuhn\footnote[2]{Spokesperson}, 
        D.~Payette, J.~Poudel, L.~Weinstein} \\
       \vspace*{0.2cm}
       \noindent \emph{Old Dominion University, Norfolk, VA 23529, USA} \\

       \vspace*{0.6cm}
       \noindent {R.~Dupr\'{e}$^\dagger$, M.~Guidal, D.~Marchand, 
        C.~Mu\~noz, S.~Niccolai, E.~Voutier} \\
       \vspace*{0.2cm}
       \noindent \emph{Institut de Physique Nucl\'eaire, CNRS-IN2P3, Univ. Paris-Sud, 
                       Universit\'e Paris-Saclay, 91406 Orsay Cedex, France} \\
      
       \vspace*{0.6cm}
       \noindent {K.~Hafidi, Z.~Yi} \\
       \vspace*{0.2cm}
       \noindent \emph{Argonne National Laboratory, Lemont, IL 60439, USA} \\

        \vspace*{0.6cm}
       \noindent {T.~Chetry, L.~El-Fassi} \\
       \vspace*{0.2cm}
       \noindent \emph{Mississippi State University, Mississippi State, MS 39762, USA} \\

       \vspace*{0.6cm}
       \noindent {N.~Baltzell, G.~Gavalian, F.~X.~Girod, S.~Stepanyan} \\
       \vspace*{0.2cm}
       \noindent \emph{Thomas Jefferson National Accelerator Facility, Newport News, VA 23606, USA} \\

       \vspace*{0.6cm}
       \noindent {I.~Albayrak, E.~Christy, A.~Nadeeshani} \\
       \vspace*{0.2cm}
       \noindent \emph{Hampton University, Hampton, VA 23669, USA} \\

       \vspace*{0.6cm}
       \noindent {N.~ Kalantarians} \\
       \vspace*{0.2cm}
       \noindent \emph{Virginia Union University, 1500 N. Lombardy St., 
        Richmond, VA 23220, USA} \\
 
       \vspace*{0.6cm}
       \noindent {C.~Ayerbe Gayoso} \\
       \vspace*{0.2cm}
       \noindent \emph{College of William and Mary, Williamsburg, VA 23187, 
        USA} \\
        
       \vspace*{0.6cm}
       \noindent {D.~Jenkins} \\
       \vspace*{0.2cm}
       \noindent \emph{Virginia Tech, Blacksburg, VA 24061, USA} \\

        \vspace*{0.7cm}
       \noindent {\Large \textbf{A CLAS12 Run-Group Addition Proposal} } \\
      \end{center}
\renewcommand*{\thefootnote}{\arabic{footnote}}

\date{\today}

\end{titlepage}
\sloppy

\titlepage

%doublespacing
%onehalfspacing
%\linenumbers
\renewcommand{\baselinestretch}{1.10}

\setcounter{page}{5}
\addcontentsline{toc}{chapter}{Abstract}

     \begin{center}
{\large\textbf{Abstract}}
    \end{center}
\vspace*{0.4cm}

The three-dimensional picture of quarks and gluons in the nucleon is set to be 
revealed through deeply virtual Compton scattering (DVCS). With the absence of 
a free neutron target, the deuterium target represents the simplest nucleus to 
be used to probe the internal 3D partonic structure of the neutron.  We propose 
here to measure the beam spin asymmetry (BSA) in incoherent neutron DVCS 
together with the approved E12-06-113 experiment (BONuS12) within the run group 
F, using the same beam time, simply with addition of beam polarization.  The 
DVCS BSA on the quasi-free neutron will be measured in a wide range of 
kinematics by tagging the scattered electron and the real photon final state 
with the spectator proton. We will also measure BSA with all final state 
particles detected including the struck neutron. The proposed measurements is 
complementary to the approved CLAS12 experiment E12-11-003, which will also 
measure the quasi-free neutron DVCS by detecting the scattered neutron, but not 
the spectator proton. Indeed, besides providing more data for neutron DVCS, this experiment 
will allow a comparison of the measurement of the BSA of neutron DVCS from the 
approved E12-11-003 with the measurements using the two methods proposed herein. This 
comparison will help to understand the impact of nuclear effects, such as the 
final state interactions (FSI) and Fermi motion on the measurement of the 
neutron DVCS.

\newpage

\tableofcontents

\chapter*{Introduction\markboth{\bf Introduction}{}}
\label{chap:intro}
\addcontentsline{toc}{chapter}{Introduction}

Inclusive deep inelastic scattering (DIS) experiments have been instrumental in 
advancing our understanding of the QCD structure of nucleons in the past. More
recently, hard exclusive experiments such as Deeply Virtual Compton Scattering (DVCS) and 
Deeply Virtual Meson Production (DVMP) have provided important new probes that 
allows us to explore both the longitudinal motion of partons and their 
transverse spatial structure in nucleons through the generalized parton 
distribution (GPD) framework.  The GPDs correspond to the coherence between 
quantum states of different (or same) helicity, longitudinal momentum, and 
transverse position.  In an impact parameter space, they can be interpreted as 
a distribution in the transverse plane of partons carrying a certain 
longitudinal momentum~\cite{Burkardt-2000,Diehl-2002,Belitsky-2002}. A crucial 
feature of GPDs is the access to the transverse position of partons which, 
combined with their longitudinal momentum, leads to the total angular momentum 
of partons~\cite{Burkardt-2005}. This information is not accessible to 
inclusive DIS which measures probability amplitudes in the longitudinal plane.  

A high luminosity facility such as Jefferson Lab offers a unique opportunity to 
map out the three-dimensional quark and gluon structure of nucleons and nuclei.
This is one of the flagship of the JLab 12 GeV scientific program, such that many
approved proposals to the JLab Program Advisory Committee (PAC) have 
focused on studies of the 3D proton or neutron structure. We propose here to 
extend the DVCS measurements of neutrons to the effective quasi-free neutrons 
obtained using the BONuS12 
experiment. Measurements on the neutron are critical in the Jefferson Lab GPD 
program for two reasons, first as neutrons offer the only access to flavor 
decomposition for GPDs, second because of the importance of the $E$ GPD to 
measure the Ji sum rule~\cite{Ji:1996ek,Leader:2013jra}.

Recent results from CLAS on incoherent DVCS from a $^4$He target have shown that
even in light nuclei, large nuclear effects can be observed~\cite{Hattawy:2018liu}.
As this recent measurement shows a deviation of $\approx 30$\% for the beam spin asymmetry (BSA)
of DVCS of bound protons in helium, one can assume that a sizable correction 
will be necessary in the neutron case as well; although reduced in the much 
less tightly bound deuterium. This could be particularly dramatic as the BSA 
expected on neutrons are rather small, making the nuclear suppression observed in helium
of a similar size as the signal expected for the neutron. We propose here to transpose the methods 
developed for BONuS12 for PDF measurements~\cite{bonus12} to the GPDs. In 
particular, we are looking how the initial state and the break-up of the 
deuterium might affect the measured BSA. The upcoming Run group F (solely composed of
the BONuS12 experiment) scheduled 
for February-April 2020, is the perfect occasion to gather data on this topic, 
which can be done with the simple addition of beam polarization to the setting 
of the run group!

\chapter{Neutron Partonic Structure}
\label{chap:physics}

\section{DVCS Formalism and Observables}

A wealth of information on the structure of hadrons lies in the correlations 
between the momentum and spatial degrees of freedom of the partons. These 
correlations can be revealed through deeply virtual Compton scattering (DVCS), 
i.e., the hard exclusive lepto-production of a real photon, which provides 
access to a three-dimensional (3-D) imaging of partons within the generalized 
parton distributions (GPDs) framework \cite{Mueller:1998fv,
Ji:1996ek,
PhysRevD.55.7114,
Radyushkin:1996nd,
PhysRevD.56.5524}. The cross section for DVCS on a spin-1/2 target can be 
parametrized in terms of four helicity conserving GPDs: $H^q$, $E^q$, 
$\tilde{H}^q$, and $\tilde{E}^q$. The GPDs $H$, $E$, $\widetilde{H}$ and 
$\widetilde{E}$ are defined for each quark flavor (q = u, d, s, ... ).  
Analogous GPDs exist for the gluons, see references 
\cite{PhysRevD.56.5524,Goeke:2001tz} for further details.  In this work, we are 
mostly concerned by the valence quark region, in which the sea quarks and the 
gluons contributions do not dominate the DVCS scattering amplitude. The GPDs 
$H$ and $\widetilde{H}$ conserve the spin of the nucleon, while $E$ and 
$\widetilde{E}$ flip it \cite{Diehl:2001pm}. The $H$ and $E$ GPDs are called 
the unpolarized GPDs as they represent the sum over the different 
configurations of the quarks' helicities, whereas $\widetilde{H}$ and 
$\widetilde{E}$ are called the polarized GPDs because they are made up of the 
difference between the orientations of the quarks' helicities.

\begin{figure}
   \centering
   \includegraphics[width=0.60\textwidth,,clip,trim=0mm 20mm 0mm 0mm 
   ]{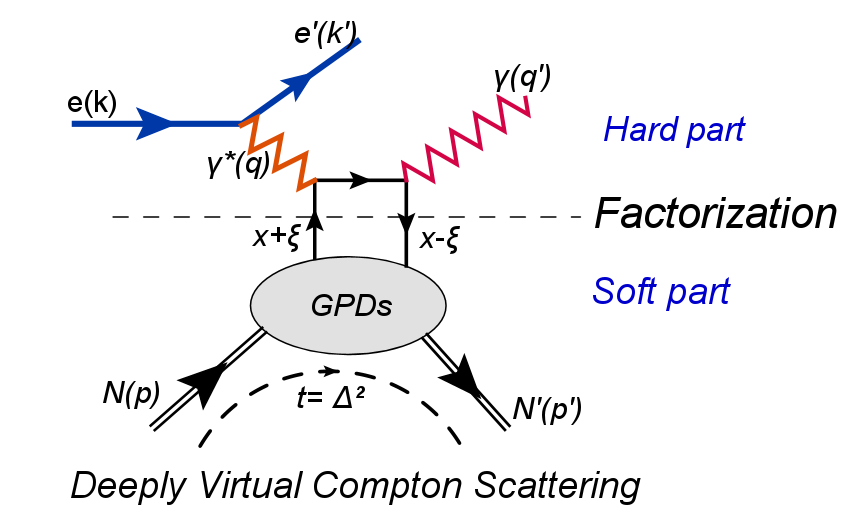}
   \caption{\label{fig:dvcshandbag} Leading-twist DVCS handbag diagram with the 
   momentum definitions labeled.}
\end{figure}

The differential cross section of leptoproduction of photons for a 
longitudinally-polarized electron beam and an unpolarized nucleon target can 
be written as:
\begin{equation}
\frac{d\sigma}{dx_B\,dy\,dt\,d\phi\,d\varphi} = \frac{\alpha^3 x_B y}{16 \pi^2 
   Q^2 \sqrt{1+\epsilon^2}} \left| \frac{\mathcal{T}}{e^3} \right|^2
\end{equation}
where $\epsilon \equiv 2x_B \frac{M_n}{Q}$, $x_B=Q^2/(2p_1\cdot q_1$) is the 
Bjorken variable, $y= (p_1\cdot q_1)/(p_1\cdot k_1)$ is the photon energy 
fraction, $\phi$ is the angle between the leptonic and hadronic planes, 
$\varphi$ is the scattered electron's azimuthal angle, $Q^2= -q_1^2$, and 
$q_1=k_1-k_2$. The particle momentum definitions are shown in 
Figure~\ref{fig:dvcshandbag}. The momentum transfer where the nucleon is 
initially at rest, $\Delta = p_1-p_2$ and $t=\Delta^2$. The Bjorken variable  
is related to another scaling variable called skewedness:
\begin{equation}
\xi = \frac{x_B}{2 - x_B} + \mathcal{O}(1/Q^2).
\end{equation}

\begin{figure}[!hbt]
   \centering
   \includegraphics[width=0.65\textwidth]{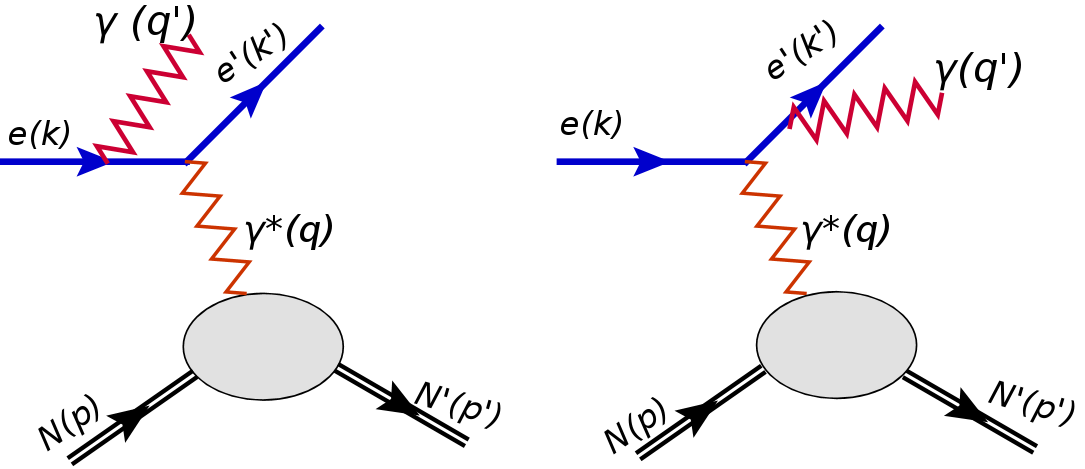}
   \caption{\label{fig:BHhandbag} BH handbag diagrams.}
\end{figure}

The amplitude is the sum of the DVCS, the Bethe-Heitler (BH), and the 
interference amplitudes, and when squared has terms
\begin{equation}
   \mathcal{T}^2 = \left|\mathcal{T}_{\text{BH}}\right|^2 + 
   \left|\mathcal{T}_{\text{DVCS}}\right|^2 + \mathcal{I}
\end{equation}
where the first is the BH contribution, the second is the DVCS part, and the last 
term is the interference part,
\begin{equation}
   \mathcal{I} = \mathcal{T}_{\text{DVCS}}\mathcal{T}_{\text{BH}}^{*} + 
   \mathcal{T}_{\text{DVCS}}^{*}\mathcal{T}_{\text{BH}}.
\end{equation}
The corresponding amplitudes are calculated with the diagrams shown in Figures 
\ref{fig:dvcshandbag} and \ref{fig:BHhandbag}. The details of contracting the 
DVCS tensor with various currents and tensors can be found 
in~\cite{Belitsky:2001ns}.
The resulting expressions for the amplitudes are
\begin{align}
   \left|\mathcal{T}_{\text{BH}}\right|^2 &= 
   \frac{e^6(1+\epsilon^2)^{-2}}{x_B^2\,y^2\,t\,
   \mathcal{P}_1(\phi)\mathcal{P}_2(\phi)} \left\{ c_0^{\text{BH}} + 
   \sum_{n=1}^{2}\left[ c_n^{\text{BH}}\cos(n\phi) +s_n^{\text{BH}}\sin(n\phi) 
   \right] \right\} \\
\left|\mathcal{T}_{\text{DVCS}}\right|^2 &= \frac{e^6}{y^2\,Q^2}\left\{ 
c_0^{\text{DVCS}} + \sum_{n=1}^{2}\left[ c_n^{\text{DVCS}}\cos(n\phi) 
   +s_n^{\text{DVCS}}\sin(n\phi) \right] \right\}\\
   \mathcal{I} &= \frac{e^6(1+\epsilon^2)^{-2}}{x_B\,y^3\,t\,
   \mathcal{P}_1(\phi)\mathcal{P}_2(\phi)}\left\{ c_0^{\mathcal{I}} + 
   \sum_{n=1}^{3}\left[ c_n^{\mathcal{I}}\cos(n\phi) 
   +s_n^{\mathcal{I}}\sin(n\phi) \right] \right\}
	\label{eq:sin}
\end{align}
The functions $c_0$, $c_n$, and $s_n$ are called \emph{Fourier coefficients}.  
They depend on the kinematic variables and the operator decomposition of the 
DVCS tensor for a target with a given spin. At leading twist there is a 
straightforward form factor decomposition which relates the vector and 
axial-vector operators with the so-called Compton form factors 
(CFFs)~\cite{Belitsky:2000gz}. The Compton form factors appearing in the DVCS 
amplitudes are integrals of the type
\begin{equation}
   \mathcal{F} = \int_{-1}^{1} dx F(\mp x,\xi,t) C^{\pm}(x,\xi)
\end{equation}
where the coefficient functions at leading order take the form
\begin{equation}
   C^{\pm}(x,\xi) = \frac{1}{x-\xi + i\epsilon} \pm \frac{1}{x+\xi - 
   i\epsilon}.
\end{equation}
We plan on measuring the beam spin asymmetry as a function of $\phi$
\begin{equation}
   A_{LU}(\phi) = \frac{d\sigma^{\uparrow}(\phi) - 
   d\sigma^{\downarrow}(\phi)}{d\sigma^{\uparrow}(\phi) + 
   d\sigma^{\downarrow}(\phi)}
\end{equation}
where the arrows indicate the electron beam helicity. 

\section{Neutron GPDs}
The measurement of free proton DVCS has been the focus of a worldwide effort 
\cite{PhysRevLett.87.182002,
   PhysRevLett.87.182001,
   PhysRevD.75.011103,
   Girod:2007aa,
   PhysRevC.92.055202,
   PhysRevLett.99.242501,
   PhysRevC.80.035206,
   PhysRevLett.114.032001,
   Jo:2015ema}
involving several accelerator facilities such as Jefferson Lab, DESY and  
CERN. These measurements now enable the extractions of GPDs and a 3-D 
tomography of the free proton \cite{Guidal:2013rya, PhysRevD.95.011501}. The 
aim of this proposal is to enhance the neutron GPD measurements along the 
approved CLAS12 experiment E12-11-003~\cite{neutronDVCS}, which will also 
measure the quasi-free neutron DVCS by detecting the scattered neutron in 
deuterium.  

In the fits of PDFs, for example \cite{Ball:2014uwa} , neutrino and deuterium 
data allow to make a flavor dependent extraction, this option is not available 
for GPDs yet due to the lack of reliable data. Indeed, the observables of DVCS, 
such as cross sections and beam-spin asymmetries are much smaller on neutron 
targets, while the nuclear effects in deuterium increase uncertainties. These 
issues have lead to results \cite{Mazouz:2007aa} which are not precise enough 
to help in a flavor dependent GPD extraction. To achieve this performance, one 
will need a large quantity of high precision data, a goal set by E-12-11-003.  
However, the impact of the uncertain initial state and the final state 
interactions on the integration of these data in global fits remain unclear.  
With this proposal, we propose to both provide more data on the neutron,
which is always important, but most importantly these data will have completely 
independent systematic uncertainties from the E-12-11-003 data. That will make 
them complementary and indicate in what ways these methods are equivalent or 
potentially need corrections.

Neutron DVCS is also hoped to provide an important contribution to the 
extraction of the GPD $E$~\cite{dHose:2016mda}. The reasoning behind this 
expectation is as follows: the GPD $E$ never appears to be dominant in the 
usually measured DVCS observables, so as a sub-leading contribution it is 
always affected with large error bars. Actually, recent extractions 
\cite{Dupre:2017hfs,Moutarde:2018kwr} show that we still barely have any 
constraint on $E$ using all the world proton data. However, as form factors 
often appear in the expressions of DVCS observables because of the Bethe 
Heitler process (see below) the situation is very different for protons and 
neutrons.  Indeed, the $F_1$ form factor of the neutron is very small, making 
the $E$ GPD more prominent in some of the neutron DVCS observables. Most 
notably the $\sin$ component of the beam spin asymmetry is defined as:

\begin{equation}
   A_{LU}^{\sin\phi} = \frac{1}{\pi} \int_{\pi}^{\pi} d\phi \sin\phi 
   A_{LU}(\phi)
\end{equation}
is proportional to the following combination of Compton form 
factors~\cite{Guidal:2013rya}

\begin{equation}
A_{LU}^{\sin\phi} \propto \Im m \left [ F_1 \mathcal{H} + \xi(F_1 + F_2) \mathcal{\tilde H} 
   - \frac{t}{4M^2} F_2  \mathcal{E} \right ],
\end{equation}
highlighting the effect of a suppressed $F_1$ on the main term.
This lead to the idea that neutron DVCS will also help significantly to constrain the 
GPD $E$. This goal is of course driven by the long standing objective of GPD physics to
measure the Ji sum rule in the nucleon:

\begin{equation}
J_q \,=\, {1 \over 2} \, \int_{-1}^{+1} d x \, x \, 
\left[ H^{q}(x,\xi,t = 0) + E^{q}(x,\xi,t = 0) \right],
\label{eq:dvcs_spin}
\end{equation}
which links the total angular momentum ($J_q$) carried by each quark $q$ to the 
sum of the second moments over $x$ of the GPDs $H$ and $E$, that will complete 
the decomposition of the nucleon spin in its various components 
\cite{Ji:1996ek,Leader:2013jra}.

\section{Measuring the neutron DVCS}

\subsection{Two new methods for two objectives}

\begin{figure}
   \centering
   \includegraphics[width=0.60\textwidth,,clip,trim=0mm 0mm 0mm 0mm 
   ]{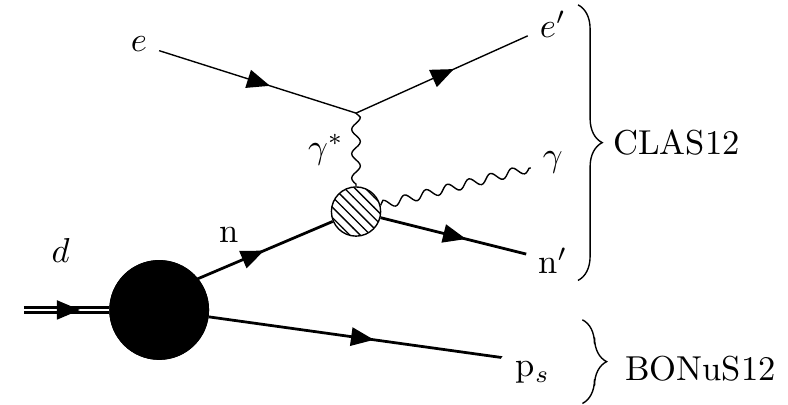}
   \caption{\label{fig:ndvcsexclusive} Fully exclusive neutron DVCS diagram in 
   deuterium. }
\end{figure}

As shown in Fig.~\ref{fig:ndvcsexclusive}, one can measure all final state 
particles of the DVCS reaction on deuterium using CLAS12 and Bonus12. This 
method, while perfect on paper, leads to an efficiency problem
as both the measurement of the spectator proton and of the neutrons are challenging
and have low efficiency (intrinsically for the neutron detection and by the 
limitation of phase space available for the proton). However, it is possible to 
ensure the exclusivity of a process when missing one of the final state 
particles by applying cuts on missing mass, momentum and energy. This strategy 
can be used, either to leave the spectator proton or the neutron undetected.  
The former approach is being used in the ongoing E12-11-003. We propose here to 
add the other option, detecting the spectator proton but not the neutron, in 
order to increase the amount of available data, but most importantly, to 
confirm the equivalency of both methods. Moreover, we will be able to
perform the fully exclusive measurement, but with rather limited statistics. We
intend to use this over-constrained last measurement to study the systematic 
effects linked to Fermi motion and final state interaction effects on our DVCS 
observable of interest, i.e., beam-spin asymmetry.

\subsection{Proton-Tagged Neutron DVCS with BONuS12}

\begin{figure}
   \centering
   \includegraphics[width=0.60\textwidth,,clip,trim=0mm 0mm 0mm 0mm 
   ]{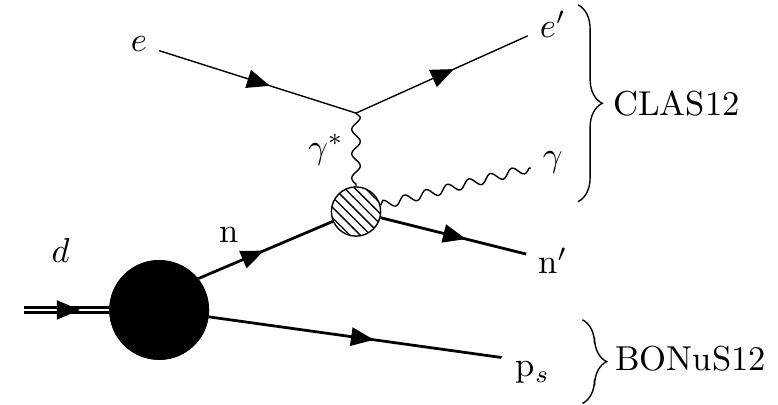}
   \caption{\label{fig:ndvcstagged} Proton-tagged neutron DVCS diagram in 
   deuterium.  }
\end{figure}

The proton-tagged neutron detection scheme is described in 
Fig.~\ref{fig:ndvcstagged}, where we see
that a detector for low energy proton spectators is necessary. Here, we propose to use the 
Bonus radial TPC, which is designed to make a similar type of measurement for inclusive 
deeply inelastic scattering. The first goal of the present proposed experiment is simply
to provide more data in the field of neutron GPD. Indeed, the measurement of neutron
DVCS is very challenging and very little published data are available at this 
point. 

Second, we observe that the systematics from this measurement are going to be 
mostly independent of the ones from E12-11-003. Indeed, while this measurement 
will be missing one of the high energy product of the reaction leading to more 
uncertainty in exclusivity cuts, it will detect the spectator proton, 
significantly helping in the understanding of the nuclear effects. This has 
interestingly an impact on both initial state and final state effects. In the 
initial state, the neutron is in fact not at rest and carries some Fermi 
momentum, which can be directly inferred from the kinematic of the spectator 
proton, thanks to the simple two body deuterium.  On the final state side, 
detecting the spectator proton in a certain range of momentum and angle allows 
to significantly reduce the probability of final state interactions to have 
occurred.

In order to resolve the initial state issue, we evaluate the standard Lorentz 
invariant $x$ and the $\gamma^{*}n$ invariant mass ($W$) in terms of the 
spectator kinematics. Both invariants, $x$ and $W$, acquire a star to indicate  
true invariants rather than the values calculated assuming a stationary, 
on-shell target:
\begin{equation}
   x^* = \frac{Q^2}{2M_{N}Ey (2-\alpha_{sp})} = \frac{x_B}{2-\alpha_{sp}},
\end{equation}
\begin{equation}
   W^* = M^{*2} - Q^2+ 2MEy(2-\alpha_{sp}),
\end{equation}

where $\alpha_{sp} = \frac{E_{s} - p^{z}_{s}}{M_N}$, with $M_N$, $E_{s}$, and 
$p^{z}_{s}$ are the on-shell mass, energy, and z-momentum component of the 
spectator proton. The off-shell mass of the bound nucleon is given by $M^{*2} = 
(M_D - E_{s})^{2} - p^{2}_{s}$, where $M_D$ is the rest mass of the deuterium.

\begin{figure}
   \centering
   \includegraphics[width=0.9\textwidth]{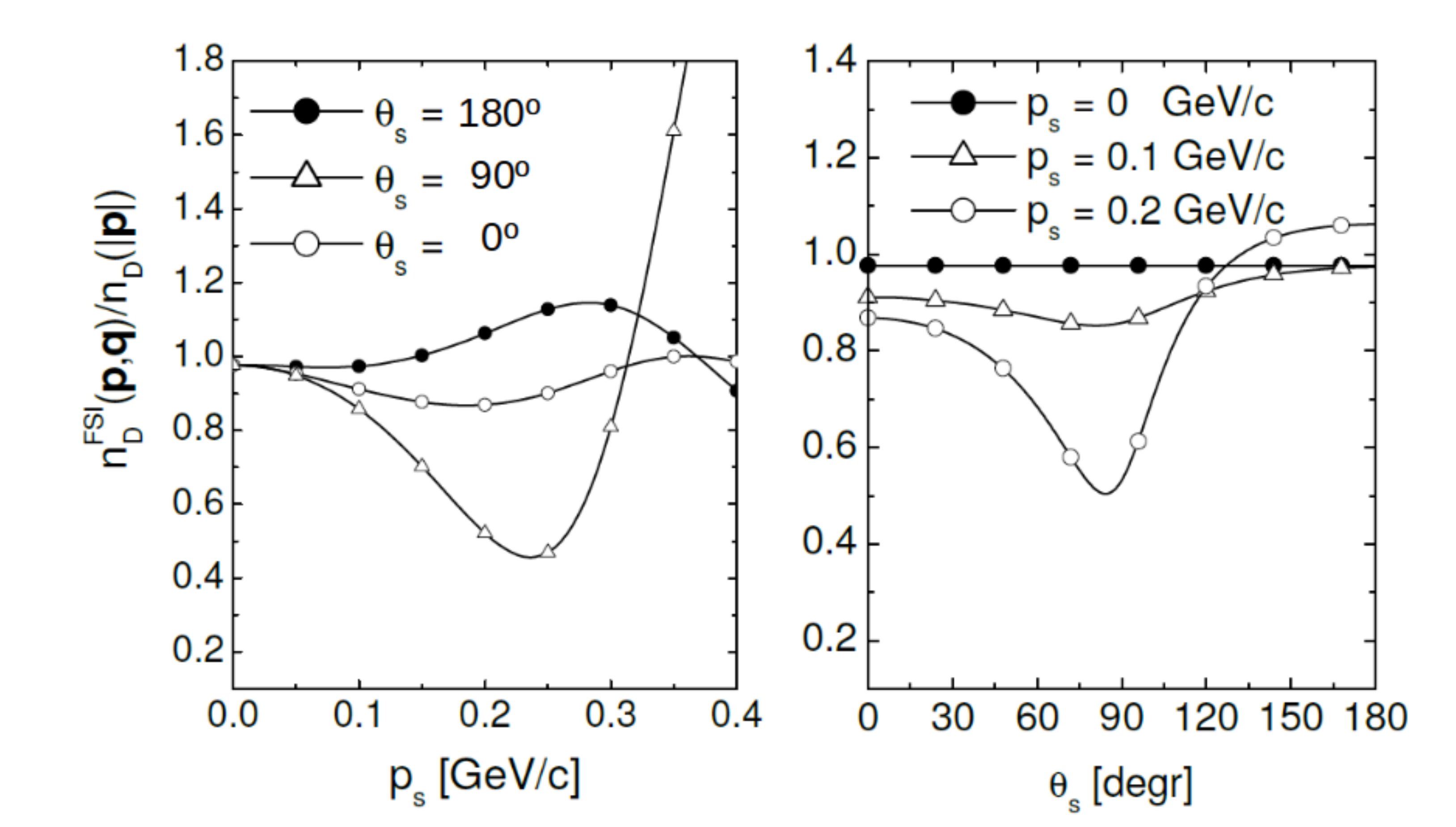}
   \caption{\label{fig:deuteronFSI} Ratio of cross sections for the FSI model 
   from~\cite{CiofidegliAtti:2003pb} to Plane Wave Impulse Approximation (PWIA) 
   calculation as a function of the spectator momentum (left) and spectator 
   angle (right). The caption on the left figure has been edited from the 
   original paper to correct for a typo.}
\end{figure}

To understand the regions where final state interactions are expected to be 
significant, we look at the spectator momentum and angle relative to the 
momentum transfer, $\theta_{s}$.  In 
Figure~\ref{fig:deuteronFSI}~\cite{CiofidegliAtti:2003pb,CiofidegliAtti:2002as}, 
we see calculations for the inclusive case.  At low recoil momentum and 
backwards spectator angle, the FSIs are negligible, whereas at high momenta 
perpendicular to the momentum transfer, FSIs are maximized.

It is important to note that the detection of the neutron or the spectator 
proton are not equivalent. While it could appear to be so after applying the 
exclusivity cut, it is not the case because of the large uncertainty (a percent 
or more) in the energy measurement of photons. This uncertainty is larger
than the momentum of the spectator proton and therefore completely hinders our 
capability to reconstruct it from missing momentum and energy methods.

\begin{figure}[tb]
\centering
\includegraphics[width=9.8cm]{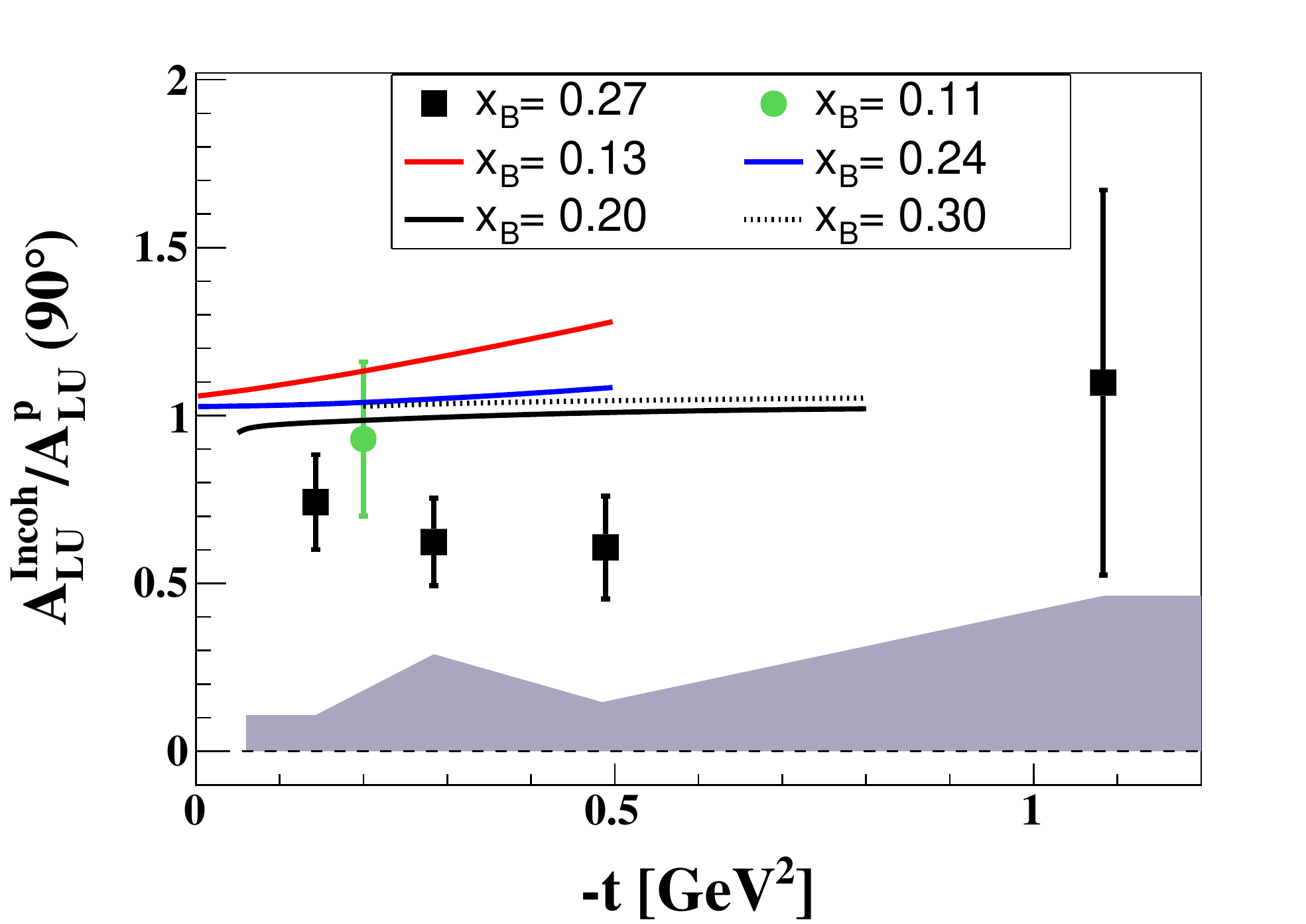}
\caption{ The $A_{LU}$ ratio of the bound to the free proton at 
   $\phi$~=~90$^{\circ}$ as a function of $t$. The black squares are from 
   CLAS-eg6 experiment~\cite{Hattawy:2018liu}, the green circle is the HERMES 
   measurement \cite{Airapetian:2009cga}.  The error bars represent the 
   statistical uncertainties, while the gray band represents the systematic 
   uncertainties. The blue and red curves are results of off-shell calculations 
   \cite{Liuti:2005gi}. The solid and dashed black curves are from on-shell 
   calculations \cite{Guzey:2008fe}.} \label{fig:incoh_EMC_ratio_ALU_proton}
\end{figure}

One should mention that our recent 
measurement~\cite{Hattawy:2018liu} of incoherent DVCS on helium conducted during the 6 
GeV era (E08-024) have shown significant modification of the proton beam 
spin asymmetry in $^4$He. These results are shown in 
Figure~\ref{fig:incoh_EMC_ratio_ALU_proton}, where we observe a much smaller asymmetry for 
bound protons. However, to this day, it is not possible to say if this suppression should
be associated to the parasitic
nuclear effects presented above or to a more fundamental source linked to the EMC effect. 
While not directly comparable, this unexpectedly large deviation seems to hint to large
nuclear effects in incoherent nuclear DVCS that have not been explored at all on the theory 
side to this day.

In conclusion, the measurement of this channel will offer a slightly smaller amount of 
data as other CLAS12 measurements. However, the key lies in the different and 
independent systematic effects of this new measurement method. In 
particular, we will study and reduce the impact of nuclear effects, which will 
significantly help in the interpretation of E12-11-003 
data~\cite{neutronDVCS}. 

\subsection{Fully exclusive neutron DVCS with BONuS12}

As highlighted above, the fully exclusive measurement of neutron DVCS is 
difficult since two hard-to-detect particles have to be measured, a slow proton 
and a fast neutron.  However, it can still provide relevant information in a 
limited kinematic space. As this measurement is fully exclusive, it provides 
over-constraints by using missing mass, momentum and energy cuts. This will 
give us a particularly clean sample of data with a minimal amount of 
corrections to be applied and the best control over systematic errors. We 
propose to use this sample, that can only be obtained using a recoil detector, 
to study the effects of Fermi motion, final state interactions and incomplete 
detection of the final state described above. This will confirm assumptions 
made in other neutron DVCS measurements and help understand better their 
systematic errors.

\chapter{Run conditions and Experimental setup}
\label{chap:physics}

\section{Run Group F conditions}

Run Group F (E12-06-113) has been approved to collect 35 PAC days (100\% 
efficiency) of data on deuterium with 11 GeV electron beam and another six days 
on hydrogen. One of the days of hydrogen data taking will be carried out with a 
low energy electron beam of about 2.2 GeV. The 40~cm long target filled with 7 
atm deuterium gas at room temperature and the 200~nA electron beam will yield a 
combined nucleon luminosity of about 2$\times$10$^{34}$cm$^{-2}$s$^{-1}$, about 
a factor of five below the standard CLAS12 nominal luminosity.

For the detection of the recoil protons, Run Group F is going to install a new 
and enlarged radial time projection chamber (RTPC) and target gas cell 
assembly, very similar to the ones used by the BONuS (E03-012) and eg6 
(E08-024) experiments. The RTPC can detect proton recoil momenta down to a 
lower limit of 70 MeV/c while being insensitive to minimum ionizing particles.  
The RTPC will be replacing the central detector's silicon tracker
and barrel micromegas, while keeping an updated design of the forward 
micromegas (FMT). In the updated version of the FMT, only three layers of 
micromegas will be kept to improve the electron's reconstructed vertex 
resolution while reducing the material in the path of the electrons. In the 
following sections we briefly introduce the experimental setup of Run Group F. 

The Run Group F approved measurement (BONuS12)~\cite{bonus12}, as well as the 
proposed measurements here, will use the same trigger as in 
RG-B~\cite{neutronDVCS}, that is an electron in Forward Tracker with road 
matching. 

%%%%%%%%%%%%%%%%%%
\section{The CLAS12 Spectrometer}
\begin{figure}
  \begin{center}
    \includegraphics[angle=0, width=0.75\textwidth]{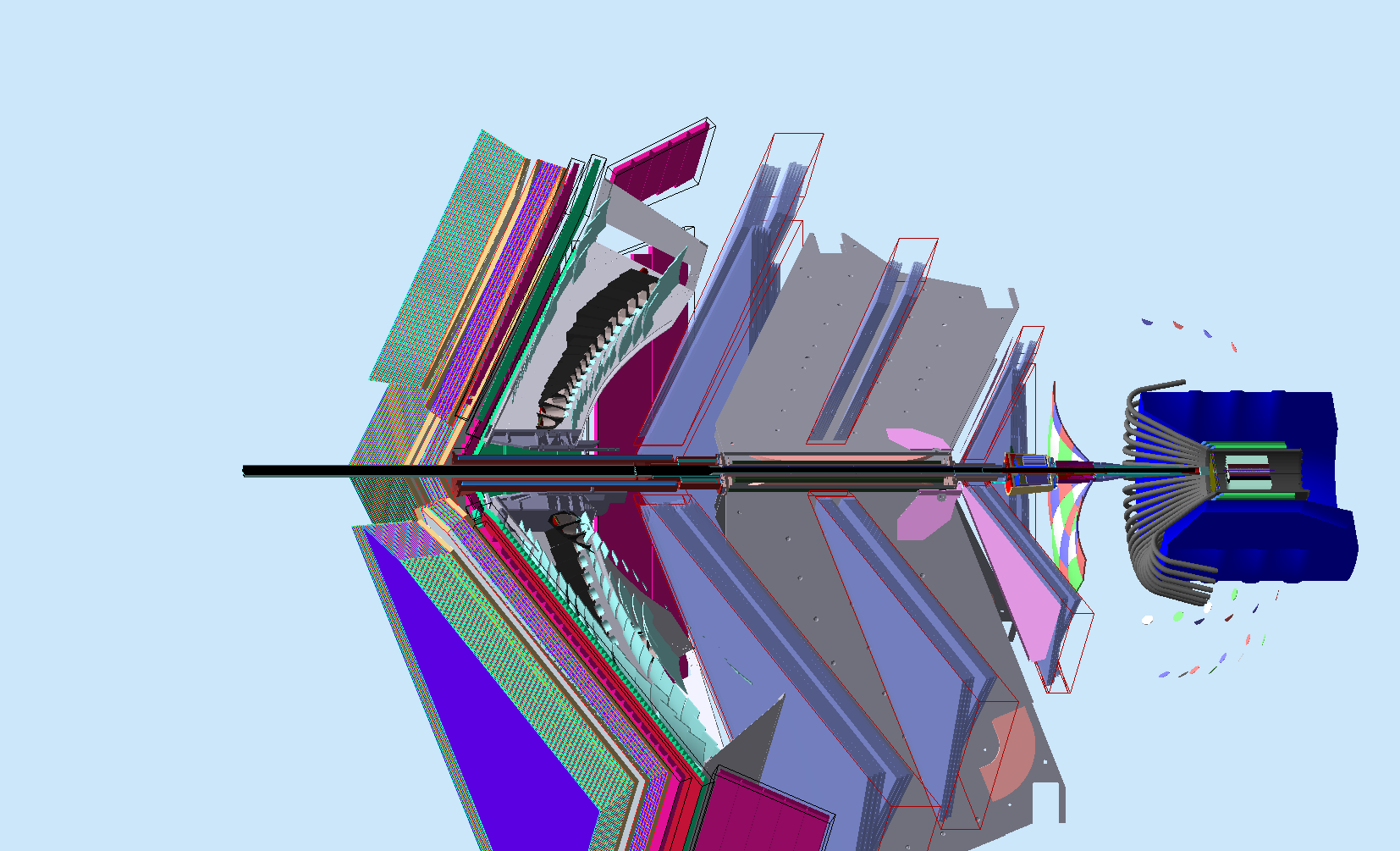}
    \includegraphics[angle=0, width=0.75\textwidth,clip, trim = 0mm 10mm 0mm 
     40mm]{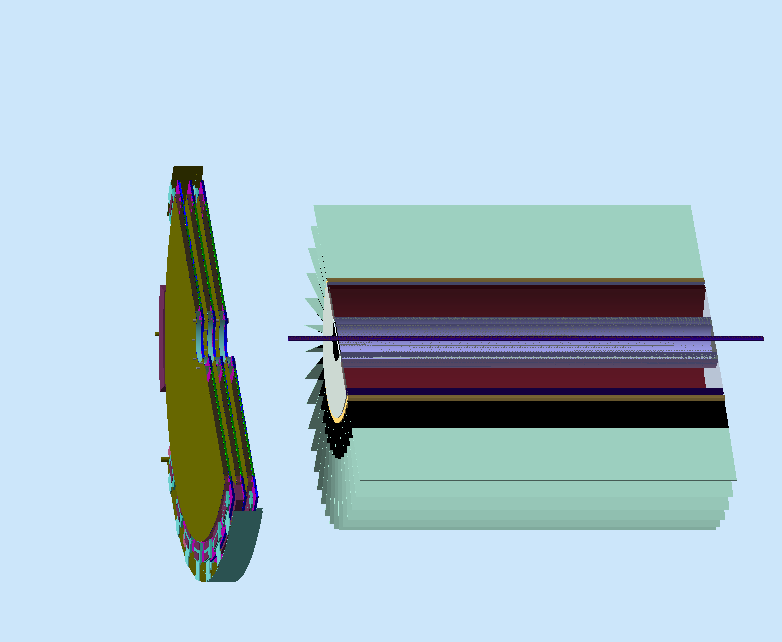}
     \caption{(Top) The schematic layout of the CLAS12 baseline design with 
     BONuS12 RTPC replacing the silicon tracker and the barrel micromegas.  
     (Bottom) Schematic layout showing BONuS12 RTPC with the modified design of 
     the forward micromegas.}
    \label{fig:fd}
  \end{center}
\end{figure}

%%%%%%%%%%%%%%%%%%
The CLAS12 spectrometer is designed to operate with 11~GeV beam at an 
electron-nucleon luminosity of $\mathcal{L} = 
1\times10^{35}~$cm$^{-2}$s$^{-1}$. The baseline configuration of the CLAS12 
detector consists of the forward detector and the central detector 
packages~\cite{CD} (see Figure~\ref{fig:fd}).
The CLAS12 Central Detector~\cite{CD} is designed to detect various charged 
particles over a wide momentum and angular range. The main detector package 
includes:
\begin{itemize}
 \item Solenoid Magnet: provides a central longitudinal magnetic field up to 
5~Tesla, which serves to curl emitted low energy M{\o}ller electrons and determine 
particle momenta through tracking in the central detector.
 \item Central Tracker: consists of 3 double layers of silicon strips and 6 
    layers of Micromegas. The thickness of a single silicon layer is  
    \SI{320}{\um}.
 \item Central Time-of-Flight (CTOF): an array of scintillator paddles with a 
    cylindrical geometry of radius 26 cm and length 50 cm; the thickness of the 
      detector is 2 cm with designed timing resolution of $\sigma_t = 50$ ps, 
      used to separate pions and protons up to 1.2 GeV/$c$.
\end{itemize}

We will use the forward detector for electron, photon, and neutron detection.  
The central detector's silicon tracker and barrel micromegas will be removed to 
leave room for the BONuS12 RTPC for the detection of the slow recoiling protons 
in deuterium. In addition to the main CTOF of CLAS12 Central detector, we will 
be using the Central Neutron Detector (CND) for the detection of the final 
state recoiling neutrons.  

The scattered electron, photon, and some neutrons will be detected in the 
forward detector which consists of the High Threshold Cherenkov Counters 
(HTCC), Drift Chambers (DC), the Low Threshold Cherenkov Counters (LTCC), the 
Time-of-Flight scintillators (TOF), the Forward Calorimeter and the Preshower 
Calorimeter. The charged particle identification in the forward detector is 
achieved by utilizing the combination of the HTCC and TOF arrays with the 
tracking information from the Drift Chambers. The HTCC together with the 
Forward Calorimeter and the Preshower Calorimeter will provide a pion rejection 
factor of more than 2000 up to a momentum of 4.9~GeV/c, and a rejection factor 
of 100 above 4.9 GeV/c. The photons and the neutrons are detected using the 
calorimeters. As will be showing later on, the majority of the final state 
recoiling neutrons will be detected using the sub-detectors of CLAS12 Central 
Detector, in particular CTOF and CND. 

\section{BONuS12 RTPC} 

\begin{figure}
  \begin{center}
    \includegraphics[angle=0, width=0.45\textwidth, clip,trim=50mm 10mm 80mm 
     0mm]{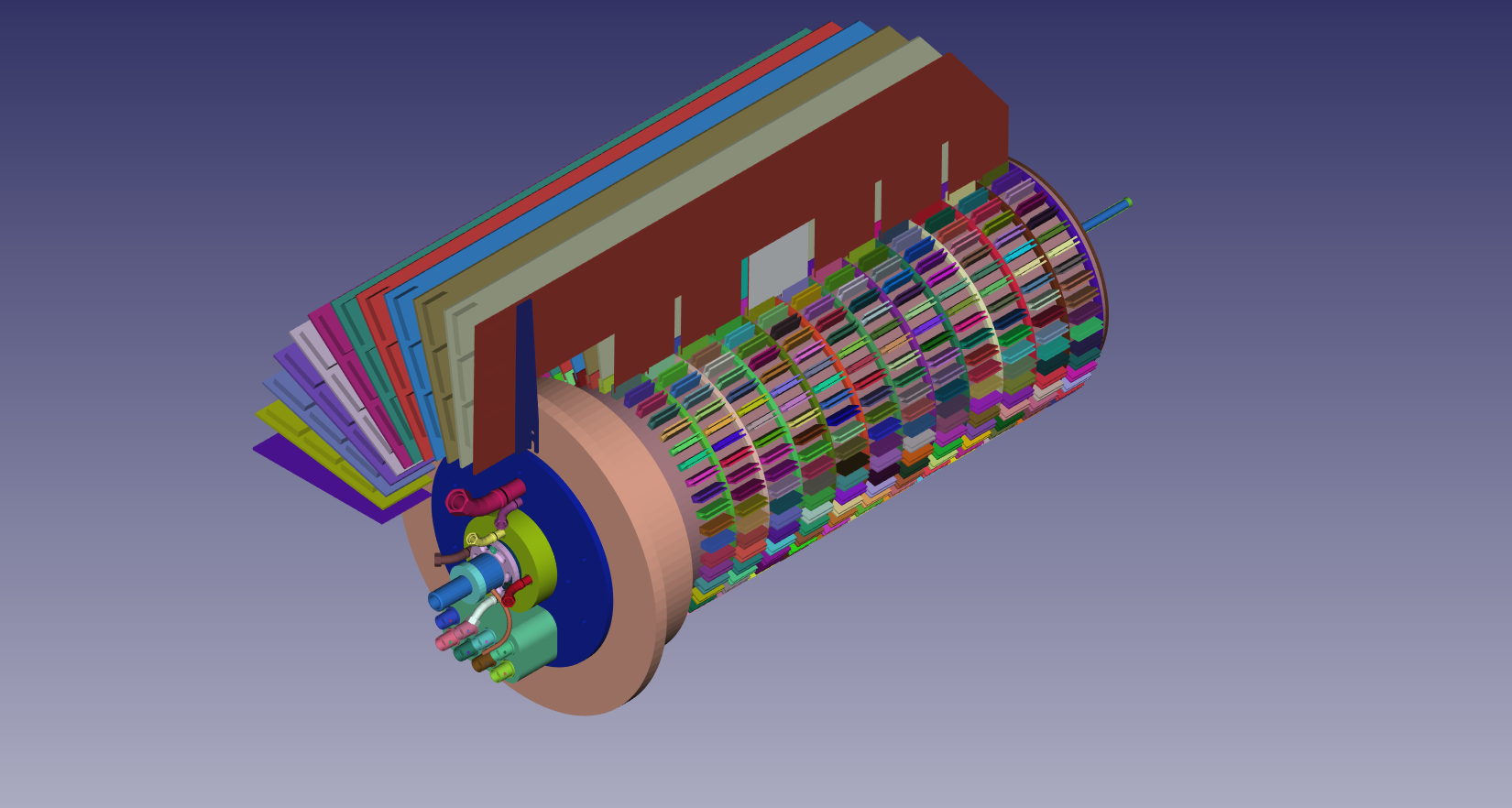}
    \includegraphics[angle=0, width=0.45\textwidth,clip,trim=0mm 10mm 20mm 0mm 
     ]{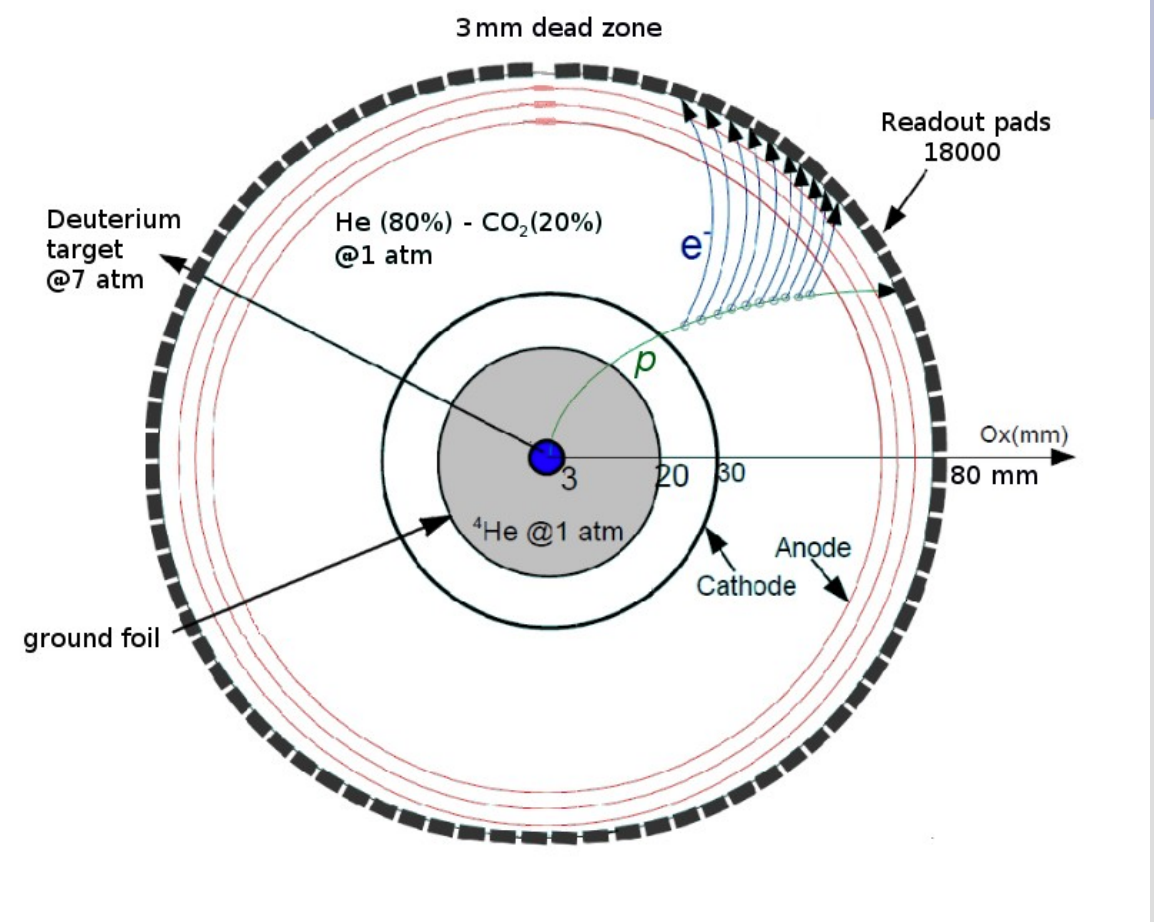}
     \caption{(Left) Schematic layout showing BONuS12 RTPC showing the readout 
     padboard and few adaptor boards in addition to the gas lines. (Right) 
     Schematic drawing of the CLAS RTPC in a plane perpendicular to the beam 
     direction. See text for description of the elements.}
    \label{fig:bonus12}
  \end{center}
\end{figure}

The new CLAS12 RTPC (BONuS12) is 400~mm long cylinder of 160~mm diameter. The 
electric field is directed perpendicularly to the beam direction, such that 
drifting electrons are pushed away from the beam line.  These electrons are 
amplified by three layers of cylindrical gas electron multipliers (GEM) and 
detected by the readout system on the external shell of the detector as 
illustrated in Figure~\ref{fig:bonus12}. The BONuS12 RTPC covers almost 100\% 
of the azimuthal angle range.

We detail here the different regions shown in Figure~\ref{fig:bonus12} starting 
from the beam line towards larger radius:\\
\begin{itemize}
  \item The 7~atm Deuterium gas target extends along the beamline forming the 
     detector central axis. It is a 6~mm diameter Kapton straw with a 50~$\mu$m 
      wall of 492~mm length such that its entrance and exit 15~$\mu$m aluminum 
      windows are placed outside of the detector volume.  The detector and the 
      target are placed in the center of the solenoid.
   \item The first gas gap covers the radial range from 3~mm to 20~mm. It is 
      filled with $^{4}$He gas at 1~atm to minimize secondary interactions from
      M\o{}ller electrons scattered by the beam. This region is surrounded by a 
      4~$\mu$m thick window made of grounded aluminized Mylar.
   \item The second gas gap region extends between 20~mm and 30~mm and is 
      filled with the gas mixture of 80$\%$ $^{4}$He and 20$\%$ CO$_2$. This 
      region is surrounded by a 4~$\mu$m thick window made of aluminized Mylar 
      set at $-4260$~V to serve as the cathode.
   \item The drift region is filled with the same $^4$He-CO$_2$ gas mixture and 
      extends from the cathode to the first GEM, 70~mm away from the beam axis.  
      The electric field in this region is perpendicular to the beam and 
      averages around 550~V/cm.
   \item The electron amplification system is composed of three GEMs located at 
      radii of 70, 73 and 76~mm. The first GEM layer is set to $\Delta 
      V=-1620$~V relative to the ground and then each subsequent layer is set 
      to a lower voltage relative to the previous to obtain a strong 
      ($\sim$1600~V/cm) electric field between the GEM foils. A 275~V bias is 
      applied across each GEM for amplification.
   \item The readout board has an internal radius of 79~mm and collects charges 
      after they have been multiplied by the GEMs. Adaptor circuit boards are 
      plugged directly on its outer side and transmit the signal to the Hitachi 
      cables connected to the BMT DAQ electronics.
\end{itemize}

\section{Beam Polarization}
For our proposed measurements we ask for the electron beam to be highly 
polarized, i.e., 85\% longitudinally polarized beam, which is the average 
achieved polarization of the beam during Run Group A and Run Group B using the 
upgraded 12 GeV experimental setup. Regarding measuring the beam polarization, 
no additional commissioning time is required. The spokespersons of Run Group F 
\cite{skuhn} has accepted to schedule time to perform M{\o}ller runs once or 
twice a week, which is enough for the proposed measurements.

\chapter{Projections for the Proposed Measurements}
\label{chap:reach}
We propose to measure the neutron's DVCS beam-spin asymmetry in the following 
two channels:
\begin{enumerate}
   \item Proton-tagged neutron DVCS: $D\,+\,\gamma^{*}  \longrightarrow 
      \gamma\,+\,(n)\,+\,p$
   \item Fully exclusive n-DVCS: $D\,+\,\gamma^{*}  \longrightarrow \gamma\,+\, 
      n \,+\,p$
\end{enumerate}

To demonstrate the experimental feasibility and to extract projections for our 
proposed measurements, we present here a simulation study using a realistic 
n-DVCS event generator and the official simulation-reconstruction chain of the 
CLAS12 spectrometer augmented with BONuS12 RTPC.

\section{Monte-Carlo Simulation}
An event generator for DVCS/BH and exclusive $\pi^0$ electroproduction on the 
neutron inside a deuterium target has been developed \cite{ahmed}. The DVCS 
amplitude is calculated according to the BKM formalism \cite{Belitsky:2001ns}, 
while the GPDs have been taken from the VGG model 
\cite{PhysRevD.60.094017,Guidal:2004nd}. The Fermi-motion distribution is 
calculated with the Paris potential \cite{PhysRevC.21.861}.

The output of the event generator was fed through the  CLAS12 official 
simulation (GEMC 4.3.0)~\cite{clas12-gemc} and reconstruction (COATJAVA 
5b.7.4)~\cite{clas12-coatjava} chain, to simulate the detectors' acceptance and 
resolutions for the following final state particles, electrons, photons, and 
neutrons, within the proposed experimental setup of Run Group F. Spectator 
protons are reconstructed with a fastMC that was developed based on the 
performance of the eg6 CLAS run.

\section{Particle Identification}

The final state of proton-tagged neutron DVCS event consists of three 
particles: an electron, a proton, and a real photon, while in addition to these 
particles a neutron is required to be detected in the fully exclusive neutron's 
DVCS events. To identify the DVCS events, we first identify the different 
particles of interest. Then, events with three and four detected final-state 
particles will be further filtered by imposing the energy-momentum conservation 
laws.

For the identification of the different final state particle we use the 
official CLAS12 Event-Builder~\cite{eventbuilder}. In the following subsections 
we detail the main requirement for the different particles. 

\subsection{Electron Identification} 
For charged particles in the forward detectors, the Event Builder first assigns 
e$^-$ (PID= 11) or e$^+$ (PID= -11) (depending on the bending direction of the 
reconstructed track in the DC) if a particle satisfies all corresponding HTCC 
and ECAL requirements, and has an associated FTOF hit:
\begin{itemize}
 \item 2.0 photoelectrons in HTCC.
 \item 60 MeV in PCAL.
 \item  5-sigma cuts on a parameterized momentum-dependent sampling fraction 
    where "sampling fraction" is ECAL visible energy deposition 
      (PCAL+Inner+Outer) divided by DC momentum.  
      
\end{itemize}

\begin{figure}[tp]
\centering
   \includegraphics[width=0.48\textwidth,clip,trim=0mm 0mm 0mm 
   20mm]{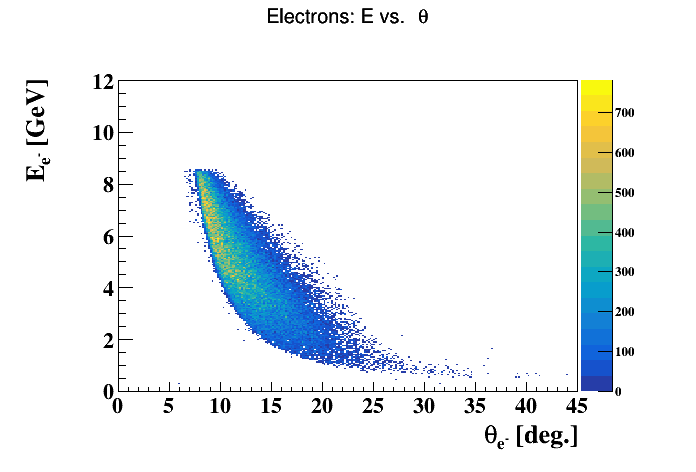}
\includegraphics[width=0.48\textwidth,clip,trim=0mm 0mm 0mm 
   20mm]{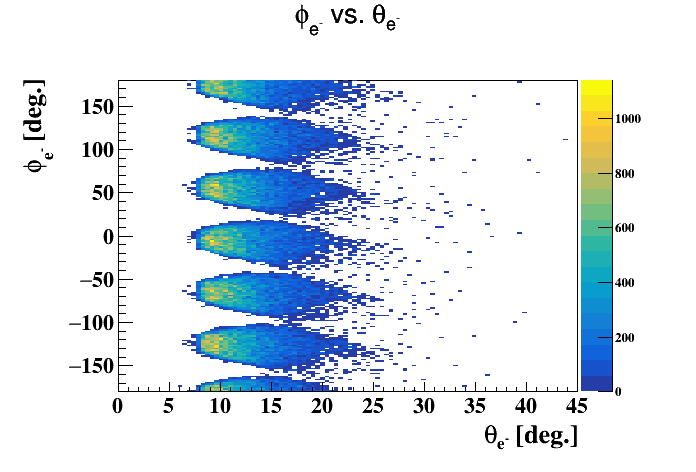}
   \caption{Electron's energy as a function of it's polar angle (left) and the 
   azimuthal angle as a function of the polar angle (right), for n-DVCS events.  
   Forward-CLAS12 acceptance and physics cuts are included.}
   \label{fig:el_kin}
\end{figure}

In addition to these initial selections cuts, some regions of CLAS12 have to be 
excluded from the analysis to ensure an accurate detection of the different 
particles. For instance, an electron that hits the edge of the EC will have 
only part of its electromagnetic shower contained within the detector. Also, 
the structure of the torus magnet divides CLAS12 into six separate sectors, 
which makes edge effects non-negligible. Figure~\ref{fig:el_kin} shows the 
kinematic distributions of the DVCS electrons being detected and reconstructed 
in the forward detector of CLAS12 spectrometer in terms of the energy as a 
function of the polar angle ($\theta$) and the azimuthal angle ($\phi$) as a 
function of $\theta$. 

\subsection{Proton Identification}
\begin{figure}[htb]
\centering
   \includegraphics[width=0.48\textwidth,clip,trim=0mm 0mm 0mm 
   20mm]{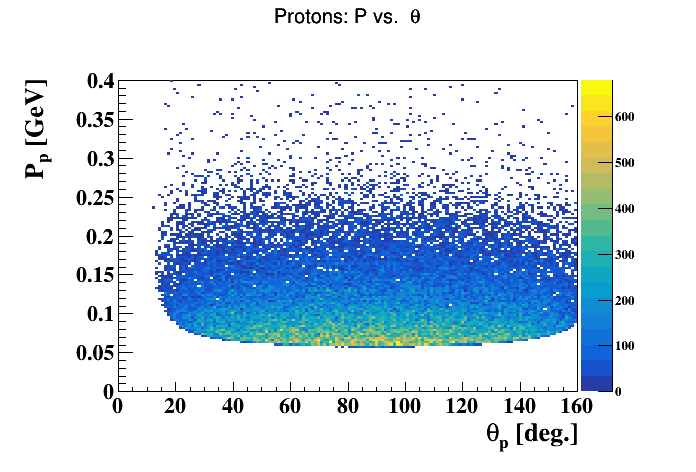}
\includegraphics[width=0.48\textwidth,clip,trim=0mm 0mm 0mm 
   20mm]{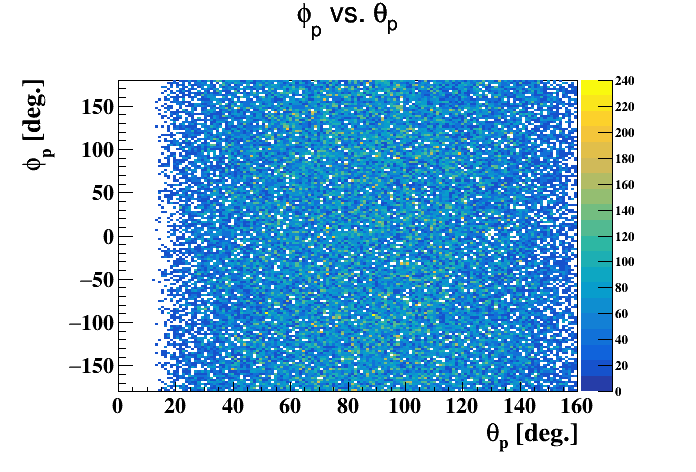}
   \caption{Recoiling proton's momentum as a function of it's polar angle 
   (left) and the azimuthal angle as a function of the polar angle (right), 
   from n-DVCS events. BONuS12 RTPC acceptance and physics cuts are included.}
   \label{fig:proton_kin}
\end{figure}

The slow recoiling final state protons will be detected within the BONuS12 
RTPC. As the RTPC is physically within the CLAS12 simulation, but the track 
reconstruction is not finalized yet within CLAS12 reconstruction, we refer to a 
realistic fastmc that reproduces the expected performance of this detector.  
This fastmc has be developed from the BONuS6 experiment and tuned very 
precisely after EG6 experiment \cite{eg6_note}, which have used very similar 
RTPCs. The geometry parameters in this fastmc have been updated with BONuS12's 
geometry. This fastmc smears the proton's kinematics and applies acceptance 
functions. Regarding the smearing, the momentum, polar angle, azimuthal angle, 
and z-vertex of the protons are smeared with Gaussians using the observed 
tracking resolutions of the RTPC (see chapter 2, section 3 in \cite{eg6_note}).  
For the acceptance, the fastmc:
\begin{itemize}
   \item ensures that the proton's track intersects with the cathode.
\item rejects the track if it goes to the upstream or the downstream detector 
   windows.
\item applies the RTPC's thresholds on the momentum and the polar angle.
\end{itemize} 
We do not specifically apply energy loss and multiple scattering in our fastmc, 
but we do apply resolution effects based on experimental data that include both 
effects. Figure~\ref{fig:proton_kin} presents the kinematic distributions of 
the recoiling final state protons within the active volume of the BONUS12 RTPC. 

\subsection{Neutrals Identification} 

\begin{figure}[htb]
\centering
   \includegraphics[width=0.48\textwidth,clip,trim=0mm 0mm 0mm 
   20mm]{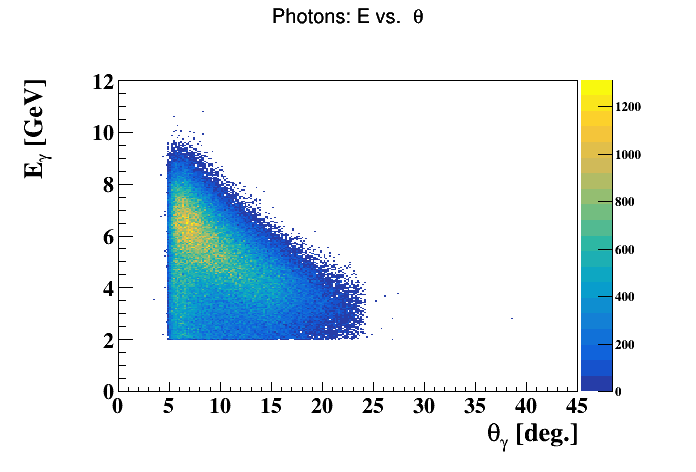}
\includegraphics[width=0.48\textwidth,clip,trim=0mm 0mm 0mm 
   20mm]{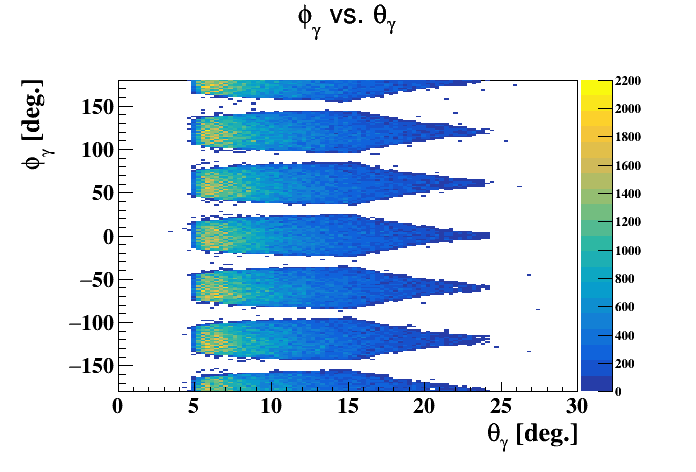}
   \caption{Photon's energy as a function of it's polar angle (left) and the 
   azimuthal angle as a function of the polar angle (right), for n-DVCS events.  
   Forward-CLAS12 acceptance and physics cuts are included.}
   \label{fig:photon_kin}
\end{figure}

\begin{figure}[htb]
\centering
  \includegraphics[width=0.48\textwidth,clip,trim=0mm 0mm 0mm 
   20mm]{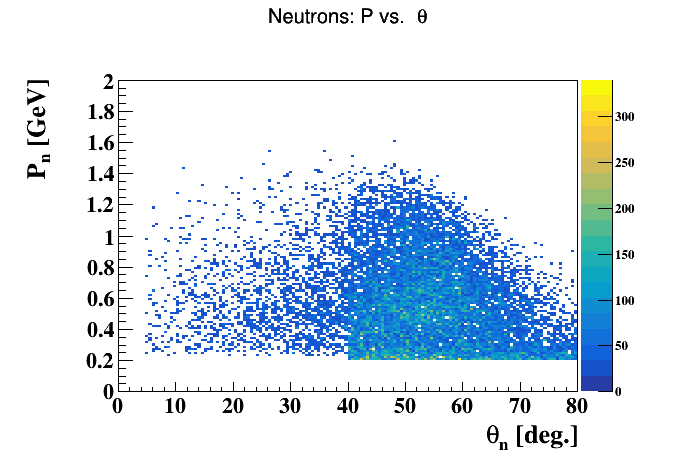}
\includegraphics[width=0.48\textwidth,clip,trim=0mm 0mm 0mm 
   20mm]{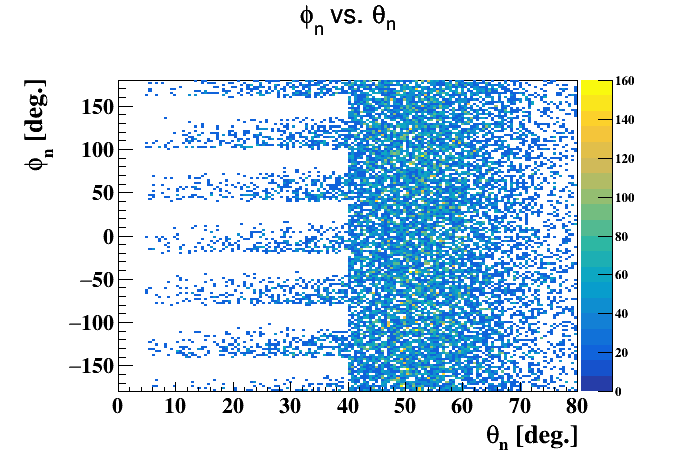}
  \caption{Recoiling neutron's momentum as a function of it's polar angle 
  (left) and the azimuthal angle as a function of the polar angle (right), from 
   n-DVCS events. BONuS12 forward and central acceptance and physics cuts are 
   included.}
  \label{fig:neutron_kin}
\end{figure}

For neutrals, only photon (PID= 22) and neutron (PID= 2112) are considered, and 
particle identification is assigned based on simple beta cut at 0.9. Currently 
only one timing response is used for this, and for ECAL the prioritization is: 
PCAL, EC Inner, EC Outer. The momentum direction is assigned based on the 
neutral's ECAL cluster position and the vertex of the charged particle used to 
determine the start time. For photons, the energy (magnitude of the momentum) 
is calculated from ECAL visible energy and momentum-dependent sampling 
fraction, while for neutrons energy is calculated from beta. For the central 
detector, CND is treated similarly as ECAL, except only neutron PID is assigned 
based on beta and the cut is at 0.8. Figure~\ref{fig:photon_kin} presents the 
kinematic distributions of the neutron-DVCS photons being detected in the 
CLAS12 forward detector. Figure~\ref{fig:neutron_kin} shows similar 
distributions of the detected final state neutrons being detected in both the 
forward and the central detectors of CLAS12 spectrometer. As can be seen from 
figures~\ref{fig:el_kin}, \ref{fig:proton_kin}, ~\ref{fig:photon_kin}, 
~\ref{fig:neutron_kin}, the DVCS electrons and photons are produced very 
forward, which is in a total agreement with our DVCS measurements during the 
6~GeV era of CLAS, while slow recoiling protons are emitted more evenly in the 
polar angle range within the acceptance of the BONuS12 RTPC. The final neutrons 
are mostly emitted above $\theta$ = 40$^{\circ}$, which is outside the 
acceptance of the forward detector of CLAS12 and was the main reason to upgrade 
CLAS12 with the Central neutron Detector during Run Group B to measure the 
neutron DVCS semi-exclusively (E12-11-003)~\cite{neutronDVCS}.

\section{Beam-Spin Asymmetry}

The beam spin asymmetry of a longitudinally polarized electron beam on an 
unpolarized target ($A_{LU}$) is defined as:
\begin{equation}
  A_{LU} = \frac{1}{P_{B}} \frac{N^{+} - N^{-}}{N^{+} + N^{-} }.
\end{equation}
where $P_{B}$ is the beam polarization, and $N^{+}$ and $N^{-}$ are the number 
of events detected with positive and negative electron helicity, respectively, 
normalized to the life-time gated Faraday cup charge for each helicity. The 
statistical uncertainty of $A_{LU}$ is
\begin{equation}
  \sigma_{A_{LU}} = \frac{1}{P_{B}} \sqrt{ \frac{1 - (P_{B}A_{LU})^{2}}{N}}
\end{equation}
where $N (= N^{+} + N^{-}) $ is the total number of measured DVCS events in 
each bin.

It is particularly convenient to use the BSA $A_{LU}$ as a DVCS observable, 
because most of the experimental systematic uncertainties, such as 
normalization and efficiencies that appear in the cross sections cancel out in 
the asymmetry ratio. However, some systematic uncertainties remain and they 
still contribute to the measured $A_{LU}$. The main known sources of systematic 
uncertainties are: the DVCS selection cuts, the fitting sensitivity to our 
binning, the beam polarization and the background (exclusive $\pi^0$) 
acceptance ratio. In the following, we present estimates of the contribution 
from each source based on our prior knowledge during CLAS-eg6 DVCS analysis 
\cite{eg6_note}.

In order to evaluate the systematic uncertainties stemming from the DVCS 
selection cuts, the eg6-analysis was repeated with changing the width of the 
exclusive cuts. The resulting systematic uncertainty to the $A_{LU}$ asymmetry
was around 6$\%$ for the incoherent DVCS channel. Because of the important 
improvement we expect with BONuS12 RTPC in terms of resolutions, we expect this 
uncertainty to be reduced to 4$\%$.

Regarding the sensitivity of the fit results to our binning, the eg6 data were 
binned into two different bin sizes in $\phi$ and the reconstructed asymmetries 
were compared. The associated systematic uncertainty for $A_{LU}$ at $\phi = 90 
^{\circ}$ was found to be 7.1$\%$. For the proposed measurements, we expect to 
achieve higher statistics and smaller bin sizes, and therefore we reduced the 
expected systematics to 3$\%$.
   
The beam polarization will be measured during the experiment by the Hall B 
M\o{}ller polarimeter. This polarimeter measures the helicity dependence of the 
M\o{}ller electron yield to obtain the beam polarization. The precision of the 
Hall B M\o{}ller polarimeter was measured to be around 3.5$\%$ 
\cite{PhysRevSTAB.7.042802}.  We assume a 3.5$\%$ systematic uncertainty on the 
measured asymmetries similar to what was achieved during the 6 GeV run.

The total systematic uncertainty is estimated to be around 11\%. To be 
conservative, in particular because expected asymmetries on neutrons are
much smaller than on protons, we used an increased total of 20\% 
systematic uncertainty for our projections. This is added quadratically to the statistical error 
bars in each bin of the reconstructed asymmetry.

\section{Projections}
Once the final state particles have been reconstructed and identified after 
passing all the physics and the geometry cuts. To consider an event as a DVCS, 
it has to pass sets of requirements: DVCS characteristic cuts and exclusivity 
cuts.

\paragraph{DVCS characteristic cuts}
\begin{itemize}
 \item $Q^{2}>$ 1 GeV$^{2}$: to ensure that the interaction occurs at the 
    partonic level and the applicability of the factorization in the DVCS  
      handbag diagram.

 \item a cut on the invariant mass ($W^*$) of the virtual photon and the target 
    neutron system to be greater than $2~GeV/c{^2}$. This cut avoids the region 
      of excitation of the neutron to resonances.
\end{itemize}

We reconstruct two types of neutron DVCS events as listed at the beginning of 
this chapter. Based on the approved  Run Group F electron-nucleon running 
luminosity, a total of 9 million tagged and 850K fully exclusive neutron DVCS 
events will be collected during the 35 PAC days. In the following two 
subsections, we present the DVCS exclusivity cuts and our proposed binning for 
each set of data selection, and finally the projections of the proposed 
measured beam-spin asymmetries.

\subsection{Proton-Tagged neutron DVCS Projections}
\begin{figure}[htb]
  \centering
    \includegraphics[width=0.95\textwidth,clip]{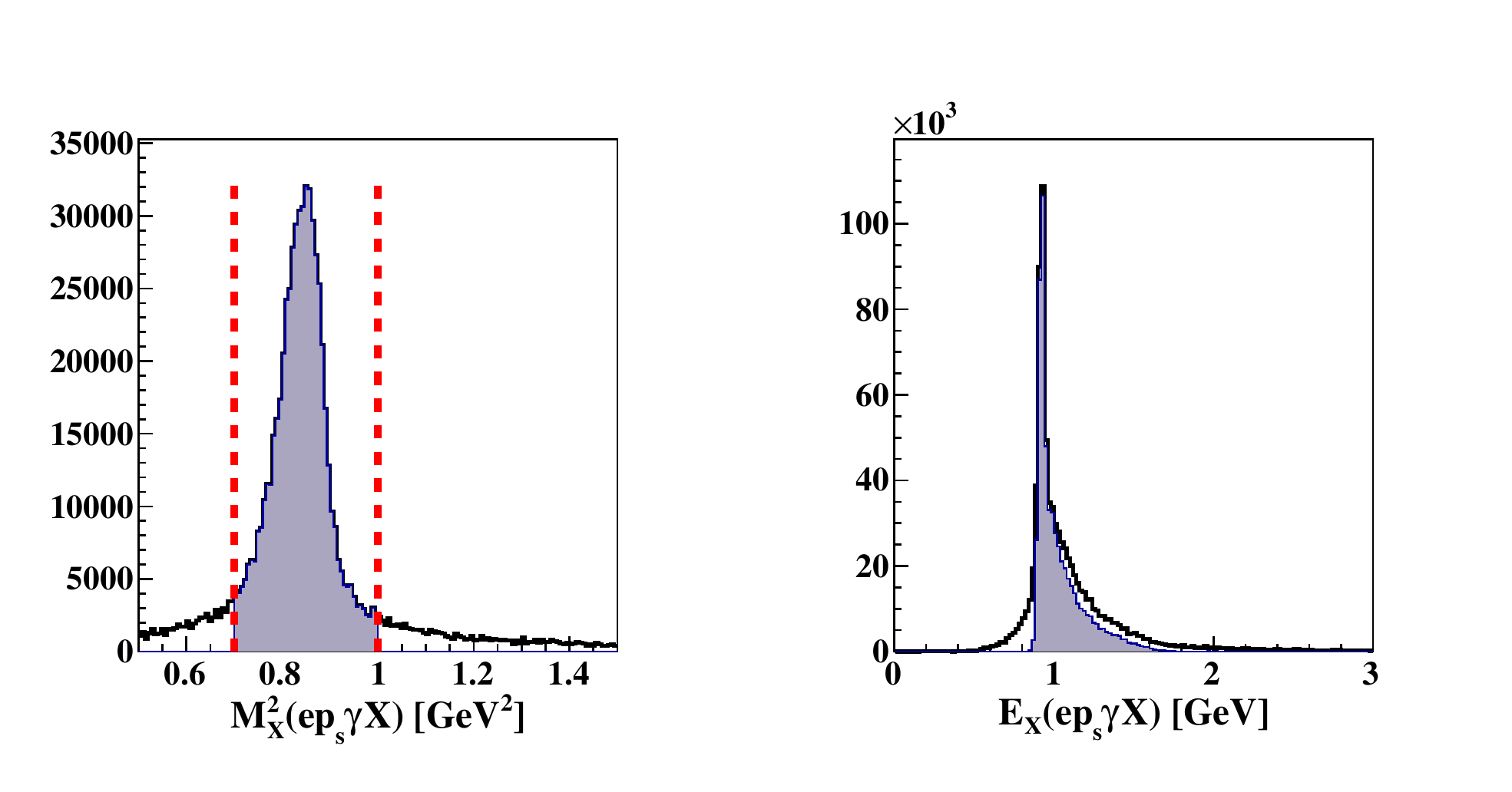}
  \caption{The distributions of the missing mass squared (left) and the missing 
   energy (right) of the identified proton-tagged neutron DVCS events. The DVCS 
   exclusivity cuts are represented by the vertical red-dashed lines. The black 
   distributions represent the incoherent DVCS event candidates before the 
   exclusivity cut. The shaded distributions represent the DVCS events that 
   passed the cut on the missing mass squared.
   \label{fig:tagged_exclusive}}
\end{figure}

After identifying the proton-tagged neutron DVCS events, i.e., having only one 
electron, one photon, and one proton in the final state, we further filter them 
by imposing the energy-momentum conservation laws.  
Figure~\ref{fig:tagged_exclusive} shows the missing mass squared distribution 
of the identified proton-tagged neutron DVCS events in addition to the missing 
energy distribution of these events. We apply an additional cut on the 
reconstructed missing mass squared to further clean the selected events as 
shown by the vertical red-dashed lines. 

\begin{figure}[htb]
  \centering
    \includegraphics[width=0.45\textwidth,clip]{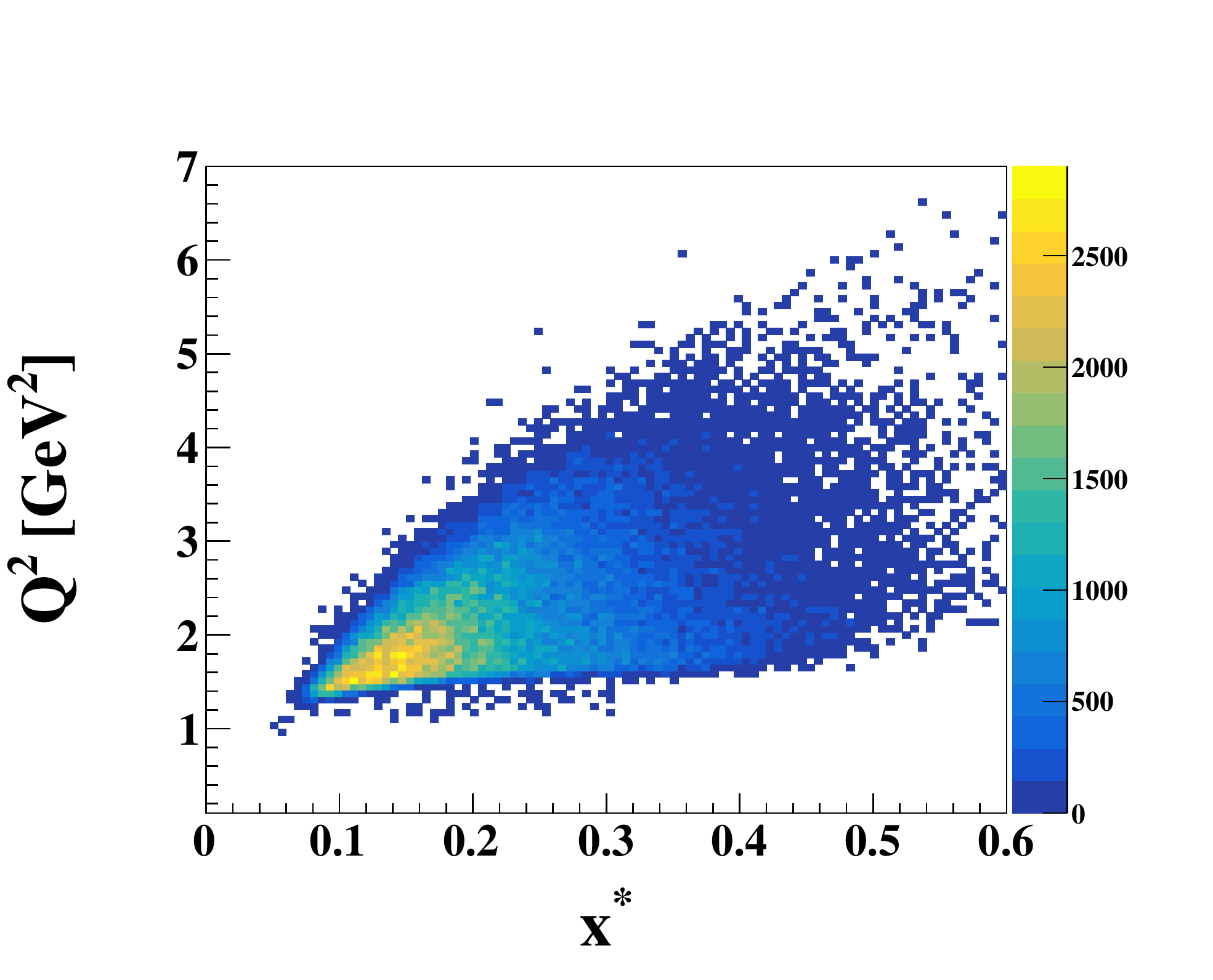}
    \includegraphics[width=0.45\textwidth,clip]{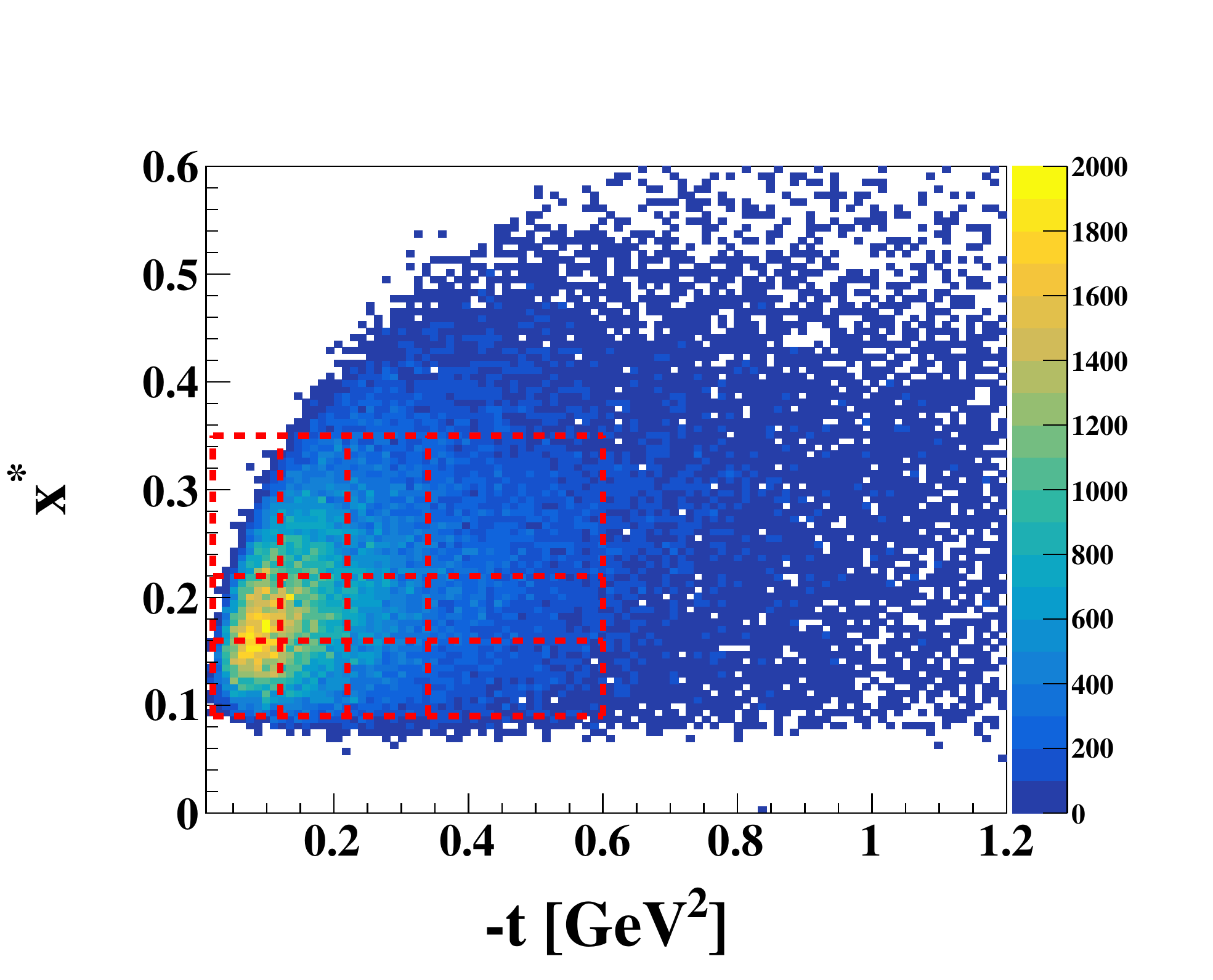}
   \caption{The distributions of the proton-tagged neutron DVCS events in terms 
   of $Q^2$ versus $x^*$ (left) and  $x^{*}$ versus $-t$ (right). On the right 
   we show the binning we propose in $x^{*}$ versus $-t$ space.
   \label{fig:binning_x_t}}
\end{figure}

The spectator approximation assumes that the recoil proton is on its mass shell 
when the electron strikes the neutron and it gains neither energy nor momentum 
during the interaction. In order to study the effect of Fermi motion on our 
measured DVCS observable, we specifically use the kinematics of the photons to 
determine the transferred momentum squared $t$.  Figure~\ref{fig:binning_x_t} 
shows the distribution of $Q^2$ as a function of $x^*$ and $x^*$ as a function 
of $t$ of the identified proton-tagged neutron DVCS events. On the right side 
plot of figure~\ref{fig:binning_x_t} we show the proposed binning in $x^{*}$ 
versus $-t$ space. The data will be binned three-dimensionally into 108 bins.  
That is, 12 bins in $x^{*}$ vs. $-t$, and then nine bins in the azimuthal angle 
($\phi$) for each of the 12 bins. The data are integrated over the full 
measured range of $Q^2$.

\begin{figure}[htb]
  \centering
    \includegraphics[width=1.1\textwidth,clip]{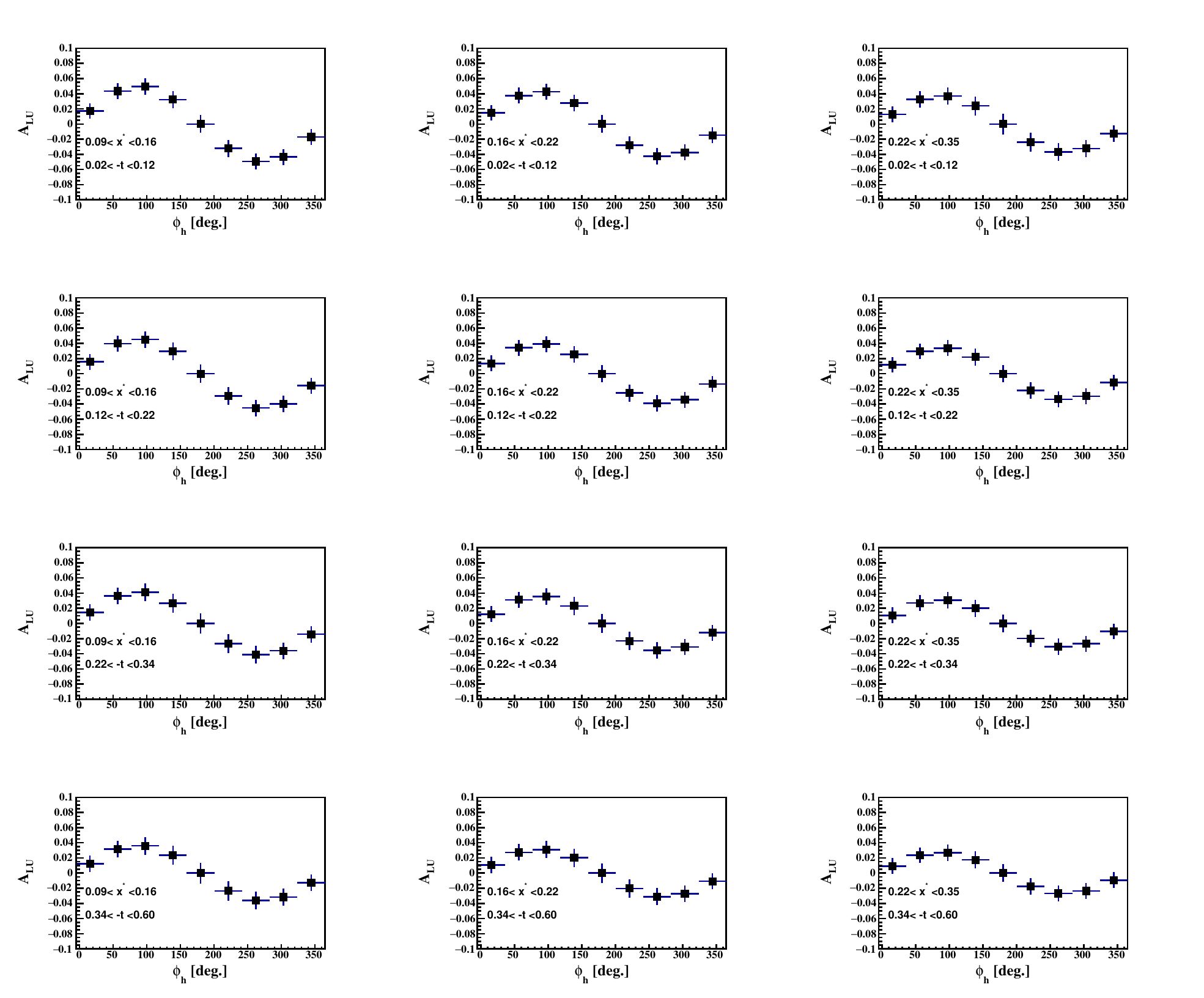}
  \caption{Projected beam-spin asymmetries as a function of the hadronic angle 
   $\phi_h$ in the 12 bins in $x^{*}$ vs $-t$ space. The error bars include 
   both the statistical and the systematic uncertainties added quadratically.
   \label{fig:alu_tagged}}
\end{figure}

Figure~\ref{fig:alu_tagged} presents the reconstructed proton-tagged neutron 
DVCS  $A_{LU}$ as a function of $\phi$ in bins of $x^{*}$ vs $-t$.  

\subsection{Fully exclusive n-DVCS projections}
Our aim is to investigate the Fermi motion and the final state interaction 
(FSI) effects on the measured neutron DVCS beam-spin asymmetry.  This can be 
achieved by comparing the measured proton-tagged neutron DVCS beam-spin 
asymmetry, that is $d(e,e'p_s\gamma)X$, to the measured full exclusive neutron 
DVCS channel, that is $d(e,e'np_s\gamma)$.  The selection of the proton-tagged 
neutron DVCS events have been presented in the previous subsection, although a 
different way of binning the data will be used here, as will be shown in this 
section. 

\begin{figure}[htb]
  \centering
    \includegraphics[width=0.95\textwidth,clip]{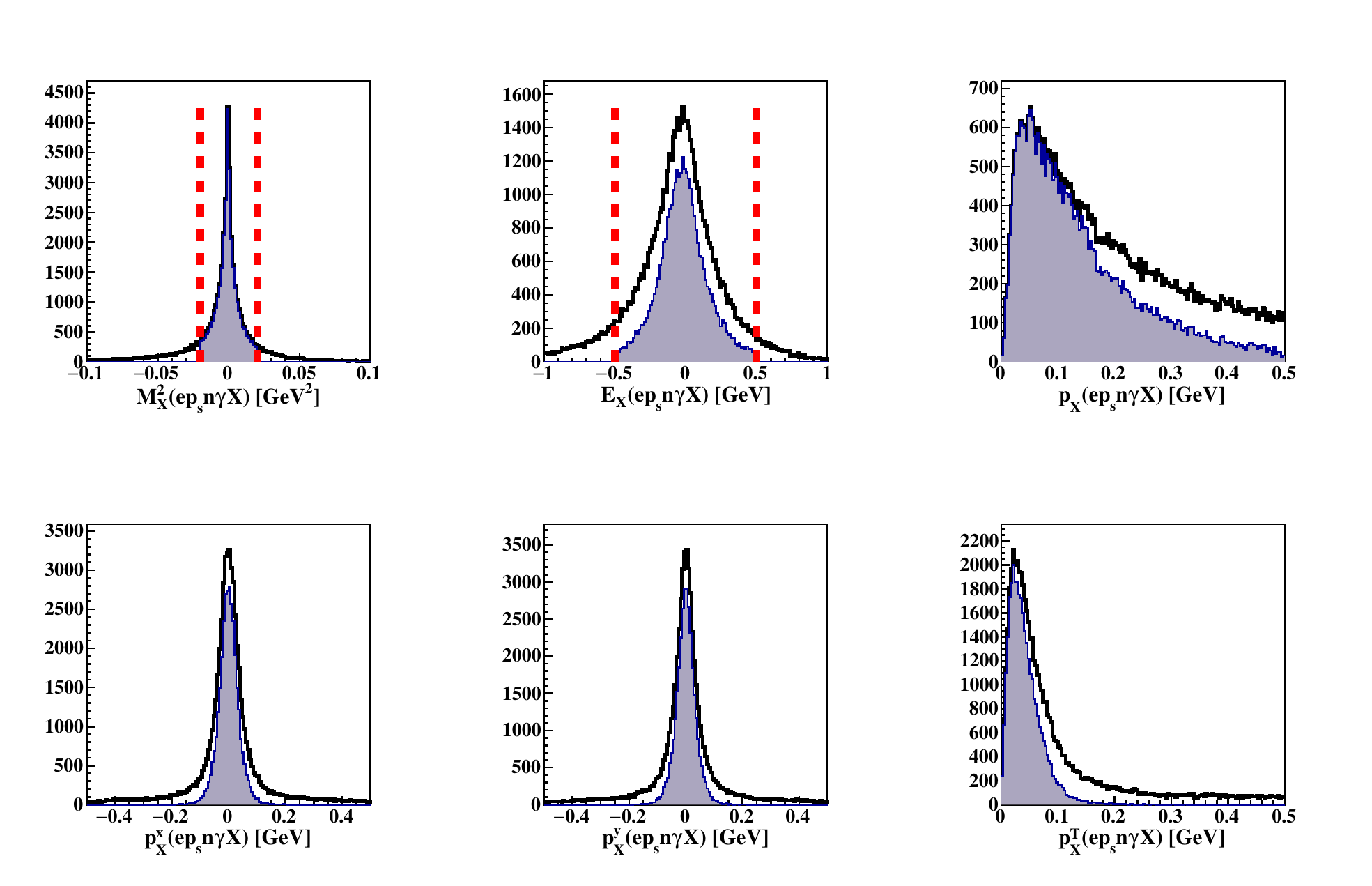}
  \caption{
    The distributions from left to right and from top to bottom are:
    missing mass squared, missing energy, missing total momentum, the 
    x-component of the missing momentum, the y-component, and the transverse 
    missing momentum in the $e'p_{s}n\gamma$ final-state system. The DVCS 
    exclusivity cuts are represented by the vertical red-dashed lines. The 
    black distributions represent the DVCS event candidates before the 
    exclusivity cuts. The shaded distributions represent the DVCS events that 
    passed all of these cuts.
   \label{fig:fully_exclusive}}
\end{figure}

Regarding the fully exclusive neutron DVCS events selection, events with the 
four final state particles will be identified after applying all the geometry 
and physics cuts on the individual final state particle. After identifying 
these events, the exclusivity of the DVCS events is ensured by imposing a set 
of constraints based on the four-momentum conservation in the reaction 
$ed\rightarrow e'p_{s}n\gamma$.  The distributions for the exclusive variables 
are shown in figure~\ref{fig:fully_exclusive}.

\begin{figure}[htb]
  \centering
    \includegraphics[width=0.45\textwidth,clip]{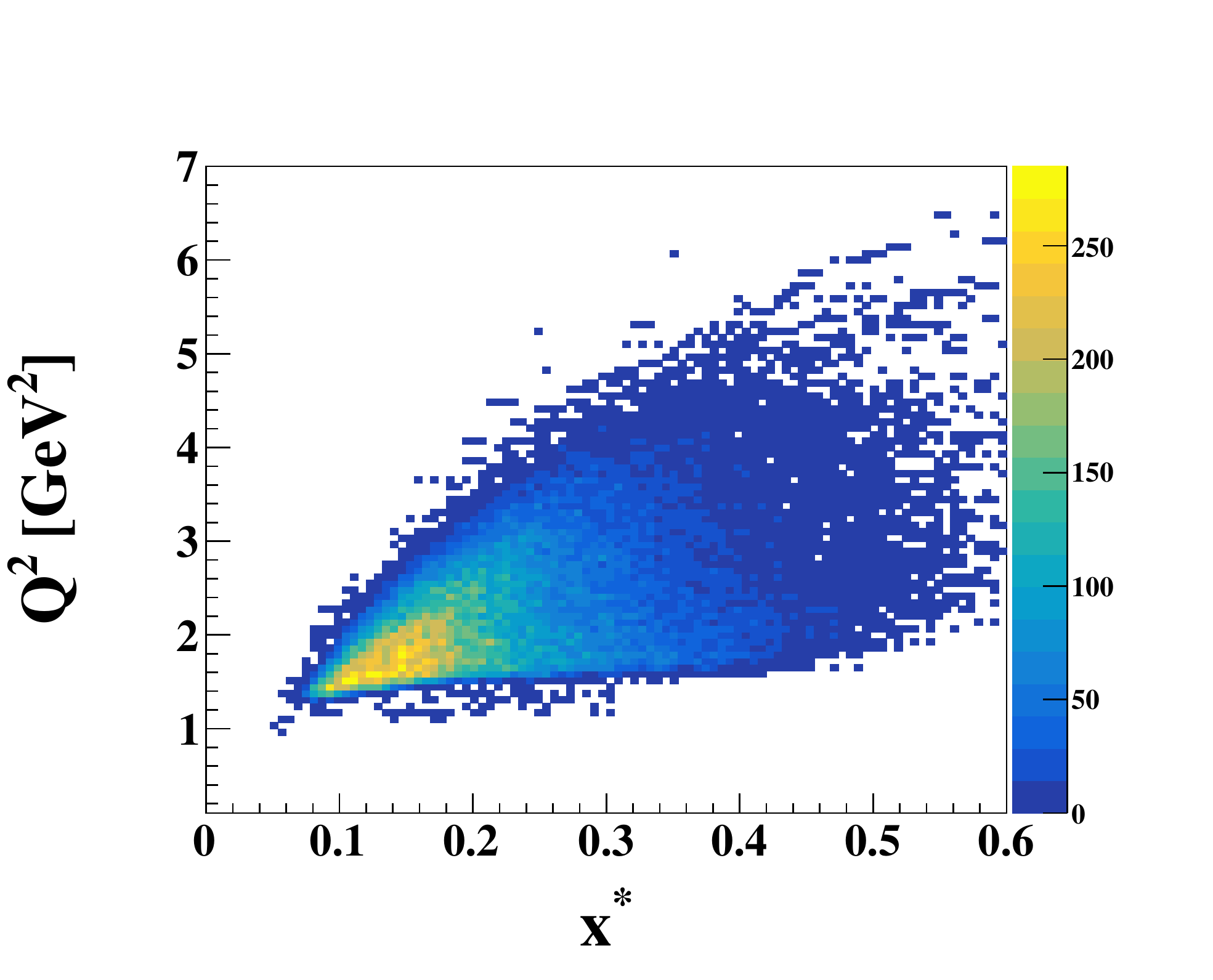}
    \includegraphics[width=0.45\textwidth,clip]{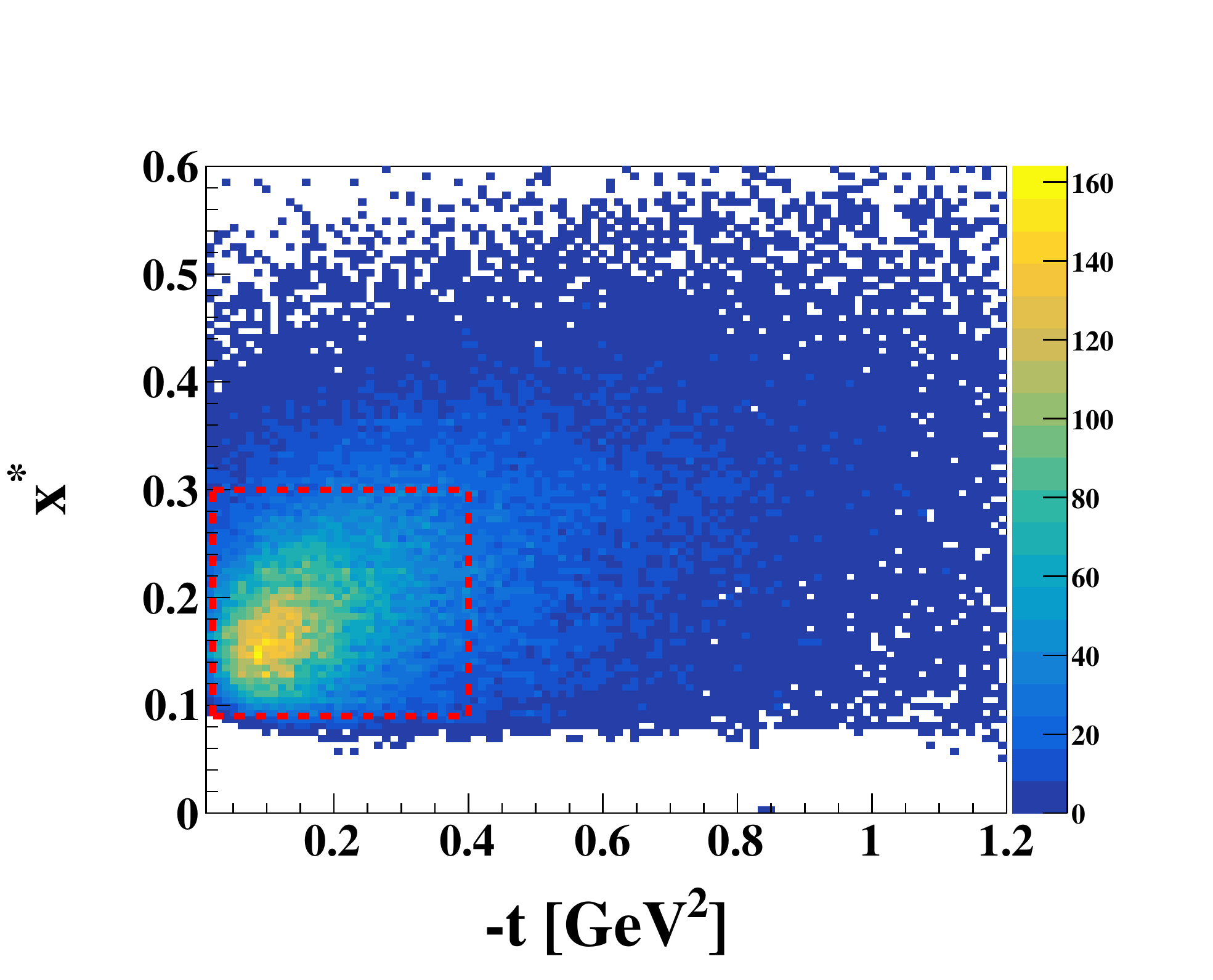}
   \caption{The distributions of the fully exclusive neutron DVCS events in 
   terms of $Q^2$ versus $x^*$ (left) and  $x^{*}$ versus $-t$ (right). On the 
   right we show the bin of interest in $x^{*}$ versus $-t$ space.
   \label{fig:exclusive_binning_x_t}}
\end{figure}

\begin{figure}[htb]
  \centering
\includegraphics[width=0.55\textwidth,clip,trim=0mm 0mm 0mm 
   20mm]{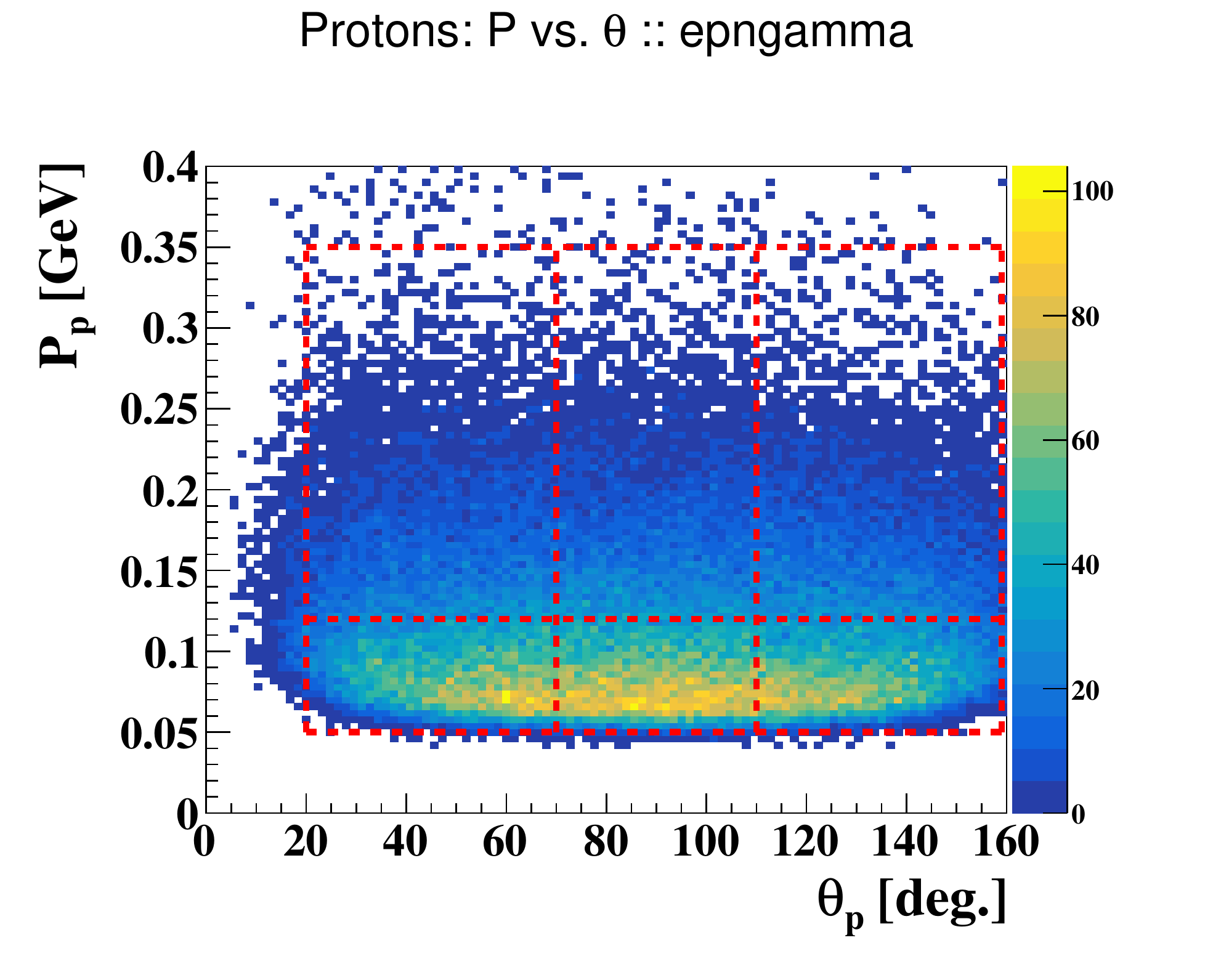}
  \caption{Data binning in $p_s$ versus $\theta_s$.
   \label{fig:ps_binning_x_t}}
\end{figure}

\begin{figure}[htb]
  \centering
    \includegraphics[width=1.1\textwidth,clip]{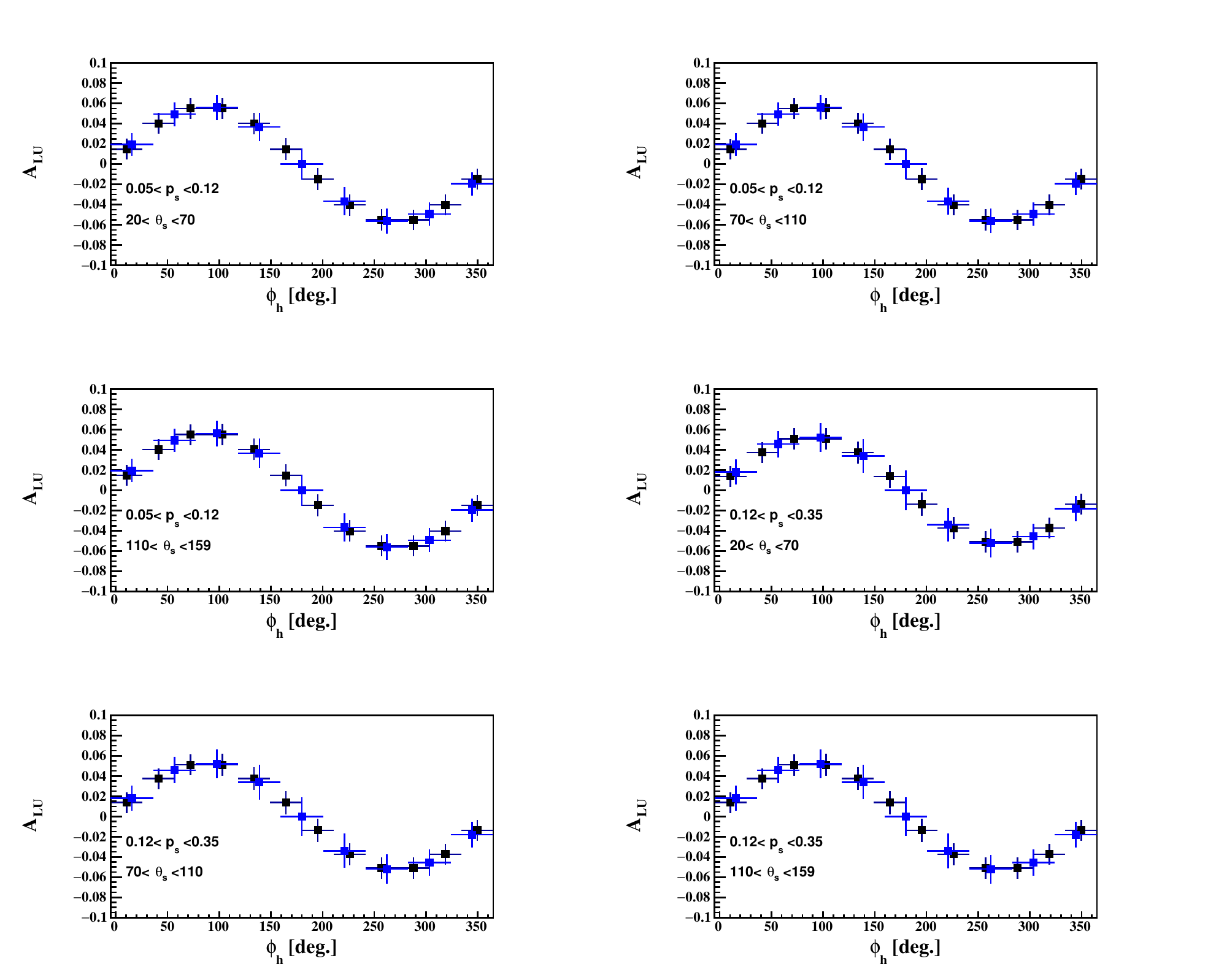}
  \caption{Projected beam-spin asymmetries as a function of the hadronic angle 
   $\phi_h$ in the binning of $p_s$ vs $\theta_s$ space for the proton-tagged 
   neutron DVCS events in black points and for the fully exclusive n-DVCS 
   events in blue points. The error bars include both the statistical and the 
   systematic uncertainties added quadratically.
   \label{fig:alu_exclusive}}
\end{figure}

As mentioned previously, a total of 850K fully exclusive events will be 
collected within the approved running luminosity and experimental setup of Run 
Group F. Similarly to the proton-tagged neutron DVCS events, we use the 
kinematics of the detected spectator proton to define the modified Lorentz 
invariant, $x^*$, and we define the transferred momentum squared, $t$, using 
the photons to investigate the initial Fermi motion effect on our DVCS 
observable of interest, $A_{LU}$. Figure~\ref{fig:exclusive_binning_x_t} shows 
the distributions of  $Q^2$ as a function of  $x^*$ and  $x^{*}$ as a function 
of $-t$ for the identified fully exclusive events. Before binning our data in 
the space of the momentum and the polar angle of the detected spectator 
momentum, we apply an initial cut on $x^{*}$ vs. $-t$, as can be seen in 
Figure~\ref{fig:exclusive_binning_x_t} which stands for the fully exclusive 
detected n-DVCS events. Similar cut is applied on the $x^{*}$ vs. $-t$ space of 
the proton-tagged neutron DVCS events. After that, both data sets are binned 
into 6 bins in the momentum of the spectator proton and its polar angle as 
shown in Figure~\ref{fig:ps_binning_x_t}. Finally, the data of each bin in 
$p_s$ versus $\theta_s$ is binned into 9 $\phi$ bins for the fully exclusive 
events and 12 $\phi$ bins for the proton-tagged neutron DVCS events. The 
reconstructed $A_{LU}$ is presented in Figure~\ref{fig:alu_exclusive} as a 
function of the hadronic angle $\phi$ for the proton-tagged neutron DVCS events 
in black points and for the fully exclusive n-DVCS events in blue points. The 
error bars in these projections include both statistical and systematic 
uncertainties. We are considering 20\% systematic uncertainties in our 
projections here as well. As one can see from our projections, the systematic 
uncertainties have the major contribution in the precision of our asymmetries 
and even with the extremely conservative assumption of 20\% consideration, we 
will observe very precise beam-spin asymmetries.        

\begin{figure}[htb]
  \centering
\includegraphics[width=0.55\textwidth,clip,trim=0mm 0mm 0mm 0mm]{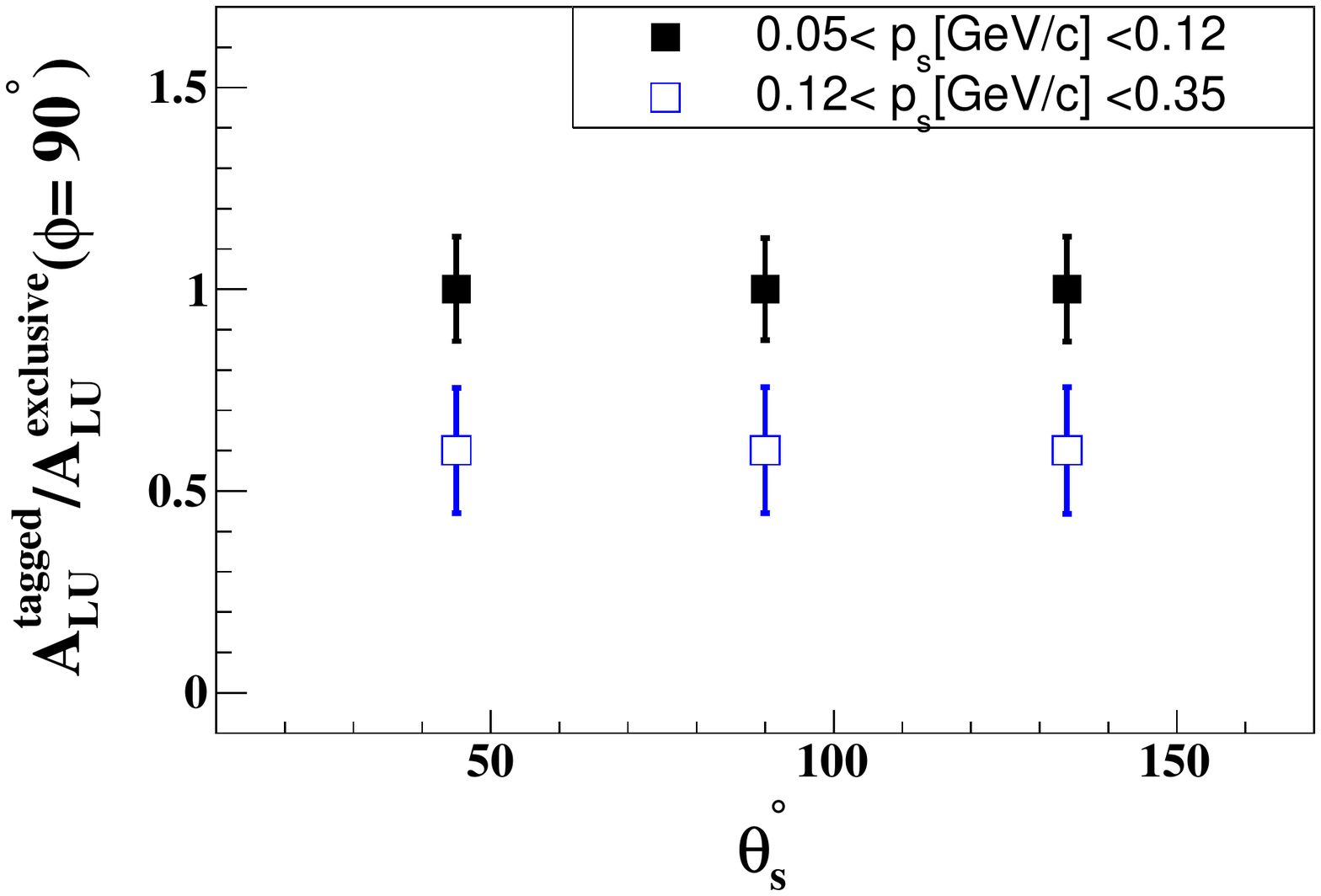}
  \caption{The projected precision on the desired measurable $A_{LU}$ ratio at 
   $\phi = 90^{\circ}$, from fitting $A_{LU}$ signals in 
   figure~\ref{fig:alu_exclusive}, of the proton-tagged neutron DVCS events to 
   the fully exclusive neutron DVCS events as a function of the spectator 
   proton $\theta_s$ in the two bins of the spectator proton momentum ($p_s$).
   \label{fig:alu_ratio}}
\end{figure}

Figure~\ref{fig:alu_ratio} presents the predicted precision on the 
reconstructed $A_{LU}$ ratio between the proton-tagged neutron DVCS events and 
the fully exclusive neutron DVCS events as a function of the spectator proton 
polar angle ($\theta_s$) in the two momentum bins shown in 
figure~\ref{fig:ps_binning_x_t}. For each data selection (tagged or exclusive)
in each bin in $p_s$ versus $\theta_s$ phase-space, each reconstructed $A_{LU}$ 
signal versus $\phi$ has been fitted with the form 
$\frac{a_{0}\sin(\phi)}{1~+~a_{1}\cos(\phi)}$. Then, $A_{LU}$ ratio is defined 
as $A_{LU}$ at $\phi = 90^{\circ}$ of the reconstructed fitted tagged DVCS 
events to the equivalent $A_{LU}$ at $\phi= 90^{\circ}$ of the reconstructed 
fitted fully exclusive DVCS events. In other words, $A_{LU}$ ratio is the 
$a_{0}$ fit parameter of the tagged DVCS to $a_{0}$ fit parameter of the fully 
exclusive DVCS beam-spin asymmetry. Such a measurement would give us insights 
about the size effect of the FSI and the Fermi motion on deep exclusive 
reactions. Since only models predicting the FSI effect on the PDFs are 
available, such valuable measurement will trigger theorists to come up with new 
models explaining our results.

 \chapter{Additional Physics Measurements with Run Group F}
 \label{chap:additional}

The combination of the high luminosity available at Jefferson Lab, the large 
acceptance of CLAS12 detector and the BONuS12 RTPC offers an amazing 
opportunity to advance our understanding of long standing questions in QCD.  
This new development in detection capabilities will allow the study of medium 
modification with a handle on Fermi motion uncertainties and FSI effects. It is 
therefore clear that the focus of the neutron DVCS proposal is only a fraction 
of the physics that can be achieved by successfully analyzing the Run Group F 
data with highly polarized beam. This data will be a gold mine, which will 
allow us to investigate in a unique way several important physics questions and 
conquer new territories in the nuclear QCD land scope. Some of the topics of 
interest to increase the physics outcome of BONuS12 polarized beam data are:

\begin{itemize}
\item Coherent DVCS and deeply virtual mesons production (DVMP) off deuteron.  
   For DVMP, we can study for example $\pi^0$, $\phi$, $\omega$ and $\rho$ 
      mesons.

\item Incoherent proton DVCS and DVMP off deuterium.

\item Deep virtual $\pi^0$ production off neutron, which is interesting by 
   itself but also the background of DVCS measurements.
\item Semi-inclusive reaction p(e,e'p)X to study the $\pi^0$ cloud of the 
   proton and $D(e, e' pp_S)X$ to study the $\pi^-$ cloud of the neutron, at 
      very low proton momenta.
   \item Transverse momentum distributions (TMDs) on the neutron (twist-3).
\item The medium modification of the transverse momentum dependent parton 
   distributions.
\item Final state interactions through the 5$^{th}$ structure function in 
$D(e,e'p_s)n$.  \end{itemize}

\chapter*{Summary \markboth{\bf Summary}{}}
\label{chap:conclusion}
\addcontentsline{toc}{chapter}{Summary}

In summary, polarizing the electron beam during the approved E12-06-113 
experiment (BONuS12) will allow us to investigate in a unique way many aspects 
of QCD within the GPD framework. The approved E12-11-003 experiment,
``Deeply Virtual Compton Scattering on the Neutron with
CLAS12 at 11 GeV'' is set to measure the n-DVCS beam-spin asymmetry by
directly detecting the struck neutron in the reaction $\gamma^{*}+d\rightarrow 
n+\gamma+(p)$. In contrast, we intend to measure the neutron DVCS beam-spin 
asymmetry by tagging the spectator slow-recoiling proton in addition to 
measuring the fully exclusive neutron DVCS channel. The first channel will 
enrich our knowledge about the partonic structure of the quasi-free neutrons, 
while the fully exclusive neutron DVCS measurement will be a golden data set to 
understand the Fermi motion and final state interaction effects on the measured 
DVCS beam-spin asymmetries. While this proposal is focusing only on the neutron 
DVCS measurements, highly polarizing the beam during Run Group F will be giving 
us a golden chance to measure additional physics topics, increasing the physics 
outcome of the approved beam time and advancing our understanding on many 
aspects of QCD.

%\begin{appendices}
%\input{review1}
%\input{review2}
%\end{appendices}

\bibliographystyle{ieeetr}
\bibliography{biblio}

\end{document}